\documentclass[aps, prb, reprint, superscriptaddress]{revtex4-2}
\usepackage{graphicx}
\usepackage{amsmath, amsfonts}
\usepackage{dsfont, mathrsfs, mathtools}
\usepackage{bm}
\usepackage{lipsum}
\usepackage{xcolor}

\DeclareSymbolFont{myletters}{OML}{ztmcm}{m}{it}
\DeclareMathSymbol{\curlylambda}{\mathord}{myletters}{"15}

\begin{document}

\title{Tower of two-dimensional scar states in a localized system}

\author{Michael Iversen}
\affiliation{Department of Physics and Astronomy, Aarhus University, DK-8000 Aarhus C, Denmark}
\author{Jens H. Bardarson}
\affiliation{Department of Physics, KTH Royal Institute of Technology, 106 91 Stockholm, Sweden}
\author{Anne E. B. Nielsen}
\affiliation{Department of Physics and Astronomy, Aarhus University, DK-8000 Aarhus C, Denmark}

\begin{abstract}
The eigenstate thermalization hypothesis describes how most isolated many-body quantum systems reach thermal equilibrium.
However, the hypothesis is violated by phenomena such as many-body localization and quantum many-body scars.
In this work, we study a finite, two-dimensional, disordered model hosting a tower of scar states.
This construction is a particular instance of a general framework and we demonstrate its generality by constructing two disordered models hosting a different tower of scar states.
At weak disorder, we find numerically that the spectra are nonthermal, and the scar states appear as exact eigenstates with high entropy for certain bipartitions. At strong disorder, the spectra localize and the scar states are identified as inverted scars since the scar states are embedded in a localized background as opposed to a thermal background. We argue that, for the considered type of models, the localization is stronger than what would be naively expected, and we show this explicitly for one of the models. The argument also provides guidelines for obtaining similarly strong localization in other scarred models.
We study the transition from the thermal phase to localization by observing the adjacent gap ratio shifting from the Wigner surmise to the Poisson distribution with increasing disorder strength.
Moreover, the entanglement entropy transitions from volume-law scaling with system size at weak disorder to area-law scaling at strong disorder.
Finally, we demonstrate that localization protects scar revivals for initial states with partial support in the scar subspace.
\end{abstract}

\maketitle
\section{Introduction}
The eigenstate thermalization hypothesis (ETH) describes how isolated many-body quantum systems reach thermal equilibrium~\cite{Deutsch1991, Srednicki1994, Rigol2008}.
The hypothesis asserts that expectation values of local observables coincide with those from the microcanonical ensemble.
ETH makes predictions about generic quantum systems and has been verified for large classes of models, see Ref.~\cite{Luca2016} and references therein.
However, several phenomena are known to violate ETH.\@

When certain interacting many-body quantum systems are exposed to strong disorder, they transition from the thermal phase to being many-body localized (MBL) \cite{Polyakov2005, Basko2006, Huse2007}.
Localized systems conflict with ETH by, e.g., being insulating at finite temperature~\cite{Prelovsek2017}, having slow entanglement growth \cite{Bardarson2012, Serbyn2013}, subsystems retaining information about the initial state after a quench \cite{Serbyn2014, Ros2017}, etc.
The nonthermal properties are attributed to the appearance of a complete set of quasi-local integrals of motion in the localized phase \cite{Abanin2013, Huse2014NovB}.
In MBL systems without a mobility edge, all energy eigenstates are nonthermal and MBL hence represents a strong violation of ETH.\@
Besides existing in disordered systems, MBL may also emerge from gradient fields \cite{Schulz2019, Nieuwenburg2019, Zhang2021} or periodic driving \cite{Bairey2017, Choi2018, Bhakuni2020}.
Numerical studies have firmly established MBL in finite systems \cite{Huse2007, Znidaric2008, Huse2010, Serbyn2014, Luitz2015, Bardarson2012, Sierant2017, Lemut2017, Clark2021} and signatures of MBL have been observed in various experimental setups \cite{Schreiber2015, Choi2016, Smidt2016, Xu2018, Rubio-Abadal2019}.
To what extent MBL remains stable in the thermodynamic limit is still being debated~\cite{Imbre2016, Suntajs2020Aug, Suntajs2020Dec, Kiefer-Emmanouilidis2021, Abanin2021, Luitz2020}.

Quantum many-body scars (QMBS) represent another violation of ETH \cite{Serbyn2021, Moudgalya2022, Chandran2023}.\@
In scarred systems, a small number of ETH-violating eigenstates are embedded in an otherwise thermal spectrum.
Hence, QMBS represents a weak violation of ETH.\@
Contrary to thermal eigenstates, scar states display sub-volume-law scaling of entanglement entropy.\@
Furthermore, when scar states are equally spaced in energy, dynamical revivals are observed from initial states in the scar subspace.
QMBS can be traced back to the discovery of analytic excited eigenstates in the Affleck-Kennedy-Lieb-Tasaki model \cite{Arovas1989} which were later recognized as scar states \cite{Moudgalya2018}.\@
Scar states have also been discovered in numerous other models \cite{Papic2018Jul, Schecter2019,Iadecola2020, Mark2020, Moudgalya2020Aug, Shibata2020, Langlett2022, Lee2020, Chertkov2021, Wildeboer2021} and several unifying formalisms have been developed \cite{Shiraishi2017, Moudgalya2020, Odea2020, Pakrouski2020, Ren2021, Moudgalya2022Sep, Ren2022, Rozon2023, Buca2023}.
QMBS have been realized in experiments with interacting Rydberg atoms \cite{Bernien2017, Bluvstein2021}, on superconducting processors \cite{Chen2022, Zhang2023, Gustafson2023}, in ultracold bosons \cite{Su2023} and in nitrogen-vacancy centers \cite{Zhou2023}.
These experiments demonstrate that, even though QMBS represent a vanishingly small fraction of the Hilbert space, they have a strong influence on system dynamics.

While MBL and QMBS are independent nonthermal phenomena, several works have attempted to realize MBL and QMBS simultaneously in one-dimensional systems \cite{Srivatsa2020, Iversen2022, Srivatsa2022, Iversen2023, Chen2023, Kolb2023}.
One approach is to exploit the analytic structure of QMBS to determine a set of local operators that annihilate the scar states.
Adding these operators with random coefficients to the Hamiltonian introduces disorder into the model without disturbing the scar states.
Disordered models with scars generally display features similar to MBL at strong disorder, e.g., energy levels following the Poisson distribution and area-law scaling of entanglement entropy with system size.
It is unclear whether these models are truly MBL in the thermodynamic limit.
But because of the similarity with MBL for finite system sizes, we refer to disordered models with scars as being localized at strong disorder.
Instead of being embedded in a thermal spectrum, the scar states reside among localized energy eigenstates.
The scar states serve as ``inverted scars'' since they are not localized and hence represent a weak violation of localization.
The interplay between localization and QMBS generates interesting effects, such as the appearance of a mobility edge \cite{Srivatsa2022} or disorder stabilization of scar revivals \cite{Iversen2023}.
However, demanding that a model hosts QMBS puts constraints on the type of disorder that can be introduced.
Consequently, these models may display a weaker type of localization and hence not localize in accordance with conventional MBL \cite{Iversen2023}.\@

In this work, we construct a two-dimensional, disordered model hosting a tower of scar states based on the rainbow scar.
At weak disorder, the majority of eigenstates near the center of each symmetry sector are thermal while the scar states are nonthermal.
When increasing the disorder strength, the model transitions from the thermal phase to being localized.
We demonstrate that the model localizes stronger than a naive prediction would suggest and that the model displays properties similar to MBL for the system sizes considered.\@
Furthermore, we present guidelines for obtaining equally strong localization in other scarred models.
The scar states display volume-law scaling of entanglement entropy for a particular bipartition while the remaining spectrum displays area-law scaling at strong disorder.
Hence, the scar states represent a tower of inverted scars which remain nonlocalized even at strong disorder.
We study the dynamics of initial states with partial support in the scar subspace and show that disorder enhances scar revivals.
Finally, we consider a different tower of scar states and construct two disordered parent Hamiltonians.
In one model, each scar state resides in a degenerate subspace for certain system sizes while in the other model, the scar states are not degenerate with other states in the spectrum.
These scar states display volume-law scaling of entanglement entropy with respect to multiple bipartitions. 
We demonstrate that the models localize at strong disorder and that the scar states represent a tower of inverted scars in both models.

In Sec.~\ref{subsec:two-dimensional-rainbow-scar}, we introduce the rainbow scar on a two-dimensional grid of spin-$1/2$ particles and discuss its basic properties.
We also describe the projections of the rainbow scar into subspaces with definite magnetization in the $z$-direction which will serve as a tower of inverted scars.
In Sec.~\ref{subsec:entanglement-entropy-of-scar-states}, we discuss the scaling of entanglement entropy with system size for the rainbow scar and the projections of the rainbow scar.
In Sec.~\ref{subsec:parent-hamiltonian}, we introduce a disordered parent Hamiltonian for the projections of the rainbow scar.
We demonstrate that the model is thermal at weak disorder and that the scar states represent nonthermal outliers.
In Sec.~\ref{subsec:introducing-example}, we characterize the localization by first introducing the relevant concepts through a simple example.
In Sec.~\ref{subsec:general-characterization}, we provide a general description of the localization and discuss guidelines for ensuring similarly strong localization in other scarred models.
In Sec.~\ref{subsec:adjacent-gap-ratio}, we verify that the model localizes by observing the level spacing statistics transition from the Wigner surmise to the Poisson distribution with increasing disorder strength.
In Sec.~\ref{subsec:entanglement-entropy}, we demonstrate that the entanglement entropy shifts from volume-law to area-law scaling with system size for increasing disorder strength as expected for localization.
In Sec.~\ref{sec:scar-dynamics}, we show that disorder stabilizes scar revivals from initial states with partial support in the scar subspace.
In Sec.~\ref{subsec:projections-of-the-rotated-EPR-scar}, we introduce a different tower of scar states.
In Sec.~\ref{subsec:entanglement-entropy-of-EPR-scar}, we demonstrate that the entanglement entropy of these scar states displays volume-law scaling with system size for multiple bipartitions.
In Sec.~\ref{subsec:parent-Hamiltonian-of-EPR-scar}, we introduce two disordered parent Hamiltonians for the tower of scar states.
In Sec.~\ref{subsec:localization-EPR}, we demonstrate that the models localize at strong disorder and that the scar states represent a tower of inverted scars in both models.
In Sec.~\ref{sec:conclusion}, we summarize our results.

\section{Model}\label{sec:model}
\subsection{Projections of the rainbow scar}\label{subsec:two-dimensional-rainbow-scar}
\begin{figure}
	\includegraphics{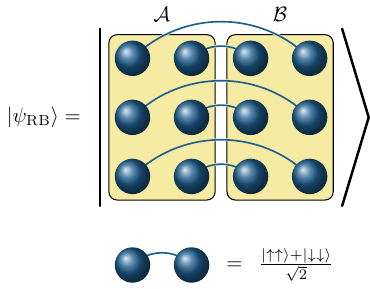}
	\caption{
		The system consists of spin-$1/2$ particles arranged in a two-dimensional grid.
		The figure illustrates a system of size $L_x \times L_y = 4 \times 3$.
		The system is separated along the vertical direction into two parts $\mathcal A$ and $\mathcal B$ of equal size.
		The rainbow scar is the tensor product of Bell states between parts $\mathcal A$ and $\mathcal B$ as described in Eq.~\eqref{eq:rainbow-scar-bell-states}.
	}\label{fig:rainbow-scar-illustration}
\end{figure}

We consider a two-dimensional grid $\bm r = (x, y)$ of size $L_x \times L_y$ occupied by spin-$1/2$ particles.
We take $L_x$ to be even throughout this work.
The system is separated into two halves $\mathcal A$ and $\mathcal B$ according to 
\begin{equation}
\begin{split}
	\mathcal A &= \Big\{(x, y) \Big| \, x = 0, \ldots, \frac{L_x}{2} - 1, \, y = 0, \ldots, L_y - 1\Big\}, \\
	\mathcal B &= \Big\{(x, y) \Big| \, x = \frac{L_x}{2}, \ldots, L_x - 1, \, y = 0, \ldots, L_y - 1\Big\}.
\end{split}
\end{equation}
We denote the Hilbert space of part $\mathcal A$ by $\mathcal H_{\mathcal A}$ and the Hilbert space of part $\mathcal B$ by $\mathcal H_{\mathcal B}$.
However, since the two parts contain the same number of sites, we have $\mathcal H_{\mathcal A} = \mathcal H_{\mathcal B}$.
Figure~\ref{fig:rainbow-scar-illustration} illustrates the system and the corresponding partitioning for system size $L_x \times L_y = 4 \times 3$.

We construct a tower of scar states based on the rainbow scar \cite{Ramirez2014, Vitagliano2010, Langlett2022}.
Let $\text{basis}(\mathcal{H}_{\mathcal{A}})$ be a basis for $\mathcal{H}_{\mathcal{A}}$.
The rainbow scar with respect to $\text{basis}(\mathcal{H}_{\mathcal{A}})$ is given by
\begin{equation}
	|\psi_\text{RB}^\text{general} \rangle = \frac{1}{\sqrt{\dim(\mathcal H_{\mathcal A})}}\sum_{|\varphi \rangle \in \mathrm{basis}(\mathcal H_{\mathcal A})} | \varphi \rangle \otimes \hat {\mathcal M} | \varphi \rangle
	\label{eq:general-rainbow-scar}
\end{equation}
where $\dim(\mathcal H_{\mathcal A})$ is the dimension of $\mathcal H_{\mathcal A}$ and $\hat {\mathcal M}$ is the mirror operator.
The mirror operator reflects each lattice site in the line $x = (L_x - 1) / 2$ without flipping the spin.
The rainbow scar has the special property that acting with any operator $\hat {\mathcal O}$ on part $\mathcal H_{\mathcal A}$ is equivalent to acting with $\hat{\mathcal M} \hat{\mathcal O}^T \hat{\mathcal M}$ on part $\mathcal H_{\mathcal B}$
\begin{equation}
	(\hat{\mathcal O} \otimes \hat{\mathds 1})| \psi_\text{RB}^\text{general} \rangle = (\hat{\mathds 1} \otimes \hat{\mathcal M} \hat{\mathcal O}^T \hat{\mathcal M})|\psi_\text{RB}^\text{general} \rangle
	\label{eq:scar-property}
\end{equation}
where $\hat{\mathds 1}$ is the identity operator and $\hat {\mathcal O}^T$ is the transpose of $\hat{\mathcal O}$ with respect to $\text{basis}(\mathcal H_{\mathcal{A}})$.
We discuss this property in Appendix \ref{appendix:scar-state-framework}.
For systems with two degrees of freedom on each site, Eq.\ \eqref{eq:general-rainbow-scar} can be rewritten as a tensor product of Bell states
\begin{equation}
	|\psi_\text{RB} \rangle = \bigotimes_{\bm r \in \mathcal A} \left( \frac{|\mathord \downarrow \rangle_{\bm r} \otimes |\mathord \downarrow \rangle_{\tilde{\bm r}} + |\mathord \uparrow \rangle_{\bm r} \otimes |\mathord\uparrow \rangle_{\tilde{\bm r}}}{\sqrt 2} \right)
	\label{eq:rainbow-scar-bell-states}
\end{equation}
where $\tilde{\bm r} = (L_x - 1 - x, y)$ is the mirror image of ${\bm r} = (x, y)$.
The rainbow scar is illustrated in Fig.\ \ref{fig:rainbow-scar-illustration}.

Following Ref.~\cite{Langlett2022}, we construct a tower of scar states from the rainbow scar by projecting $|\psi_\text{RB}\rangle$ into subspaces with definite magnetization in the $z$-direction.
Let $\hat P_M$ be the projection onto the subspace of the Hilbert space with magnetization $M$.
The normalized projection of the rainbow scar is given by
\begin{equation}\label{eq:scar-states}
	| \psi_\text{RB}^{M} \rangle = \frac{\hat P_M | \psi_\text{RB} \rangle}{\sqrt{\langle \psi_\text{RB}| \hat P_M | \psi_\text{RB} \rangle}}.
\end{equation}
Notice that the rainbow scar only has support in every other subspace, i.e., $| \psi_\text{RB}^{M} \rangle = 0$ for $M = -L_x L_y / 2 + 1, - L_x L_y / 2 + 3, \ldots, L_x L_y / 2 - 1$.
Hence, we focus on the set of scar states
\begin{align}
	\Big\{|\psi_\text{RB}^M\rangle \Big| M=-\frac{L_x L_y}{2}, -\frac{L_x L_y}{2} + 2, \ldots, \frac{L_x L_y}{2}\Big\}.
	\label{eq:projections-rainbow-scar}
\end{align}
Alternatively, these states may be constructed from the anchor state $|\Omega \rangle = \otimes_{\bm r\in\mathcal A \cup \mathcal B} |\mathord \downarrow\rangle_{\bm r}$ by acting repeatedly with the operator $\hat Q^\dagger = \sum_{\bm r\in\mathcal A} \hat S^+_{\bm r} \hat S^+_{\tilde {\bm r}}$ where $\hat S^+_{\bm r} = \hat{S}^x_{\bm{r}} + i \hat{S}^y_{\bm{r}}$, i.e., $|\psi_\text{RB}^M\rangle \propto (\hat{Q}^\dagger)^{M / 2 + L_x L_y / 4}|\Omega\rangle$.

\subsection{Entanglement entropy of the scar states}\label{subsec:entanglement-entropy-of-scar-states}
Consider separating the system into two parts $A$ and $B$ (not necessarily equal to $\mathcal{A}$ and $\mathcal{B}$).
For a state $|\psi\rangle$, the reduced density matrix is given by $\rho_{A} = \mathrm{Tr}_B(|\psi \rangle \langle \psi |)$ where $\mathrm{Tr}_B$ is the partial trace over part $B$.
The von Neumann entanglement entropy is given by $S = - \mathrm{Tr}[\rho_{A}\ln(\rho_{A})]$.
The scaling of entanglement entropy with system size of the states $|\psi_\text{RB}\rangle$ and $|\psi_\text{RB}^M\rangle$ depends on the choice of partitioning.
First, consider the vertical bipartition that separates the system into the left part $A = {\mathcal A}$ and the right part $B = {\mathcal B}$.
In this case, parts $A$ and $B$ contain one spin-$1/2$ particle from each Bell pair.
Therefore, the entanglement entropy of $|\psi_\text{RB}\rangle$ is $S = L_x L_y \ln(2) / 2$ and hence displays volume-law scaling with system size.
For the projections of the rainbow scar $|\psi_\text{RB}^M\rangle$, the entanglement entropy also displays volume-law scaling with system size \cite{Langlett2022}.
Next, let $L_y$ be even and consider the horizontal bipartition where $A$ consists of the spin-$1/2$ particles in the top half and $B$ consists of the bottom half.
In this case, each Bell pair is fully contained in either $A$ or $B$ and the entropy of the rainbow scar vanishes $S = 0$.
The entropy of the states $|\psi_\text{RB}^M\rangle$ is identical to the entropy of the ``fine-tuned'' cut in the corresponding one-dimensional model \cite{Langlett2022}.
Therefore, the entropy of $|\psi_\text{RB}^M\rangle$ displays logarithmic scaling with system size for the horizontal bipartition.

\subsection{Parent Hamiltonian}
\label{subsec:parent-hamiltonian}
\begin{figure*}
	\centering
	\includegraphics{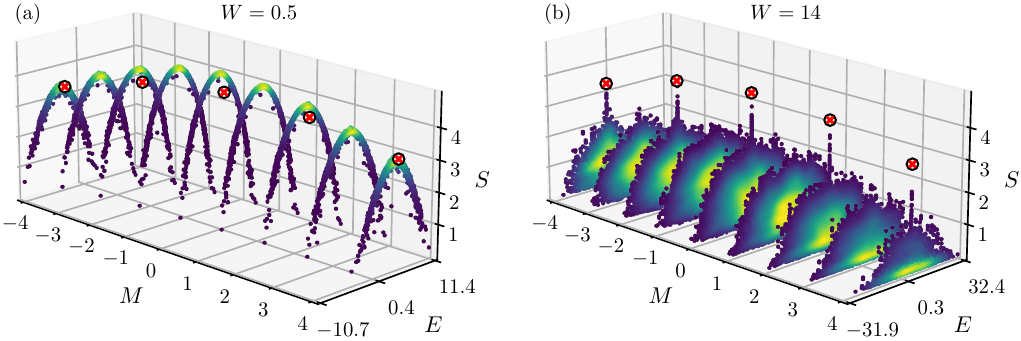}
	\caption{
	The entanglement entropy of eigenstates of the Hamiltonian from Eq.~\eqref{eq:general-Hamiltonian} as a function of energy for several magnetization sectors.
	The entanglement entropy is computed with respect to the bipartition $A = \mathcal A$ and $B =\mathcal B$ and each panel displays the results for a single disorder realization.
	We consider system size $L_x \times L_y = 4 \times 4$, parameter values $J = c = \mu = 1$ and (a) weak disorder $W = 0.5$ and (b) strong disorder $W = 14$.
	The color intensity illustrates the density of points and lighter (darker) colors display higher (lower) density of points.
	The scar states are shown as crosses enclosed by a circle.
	}
	\label{fig:entropy-vs-disorder}
\end{figure*}
Consider a general Hamiltonian of the form
\begin{equation}
	\hat H = \hat H_{\mathcal A} \otimes \hat {\mathds 1} + \hat {\mathds 1} \otimes \hat H_{\mathcal B} + \hat H_{\mathcal A \mathcal B} + \hat H_\text{SG}.
	\label{eq:general-Hamiltonian}
\end{equation}
where $\hat H_{\mathcal A}$ acts within $\mathcal H_{\mathcal A}$ and $\hat H_{\mathcal B}$ acts within $\mathcal H_{\mathcal B}$.
The operators $\hat H_{\mathcal A \mathcal B}$ and $\hat H_\text{SG}$ act on degrees of freedom in both $\mathcal H_{\mathcal A}$ and $\mathcal H_{\mathcal B}$.
We choose the first term according to
\begin{equation}
	\hat H_{\mathcal A} = 
	J \sum_{\substack{\bm r, \bm r' \in \mathcal A \\ \langle \bm r, \bm r' \rangle}} \bm S_{\bm r} \cdot \bm S_{\bm r'} 
	+ \sum_{\bm r \in \mathcal A} h_{\bm r} \hat S_{\bm r}^z
	\label{eq:HA}
\end{equation}
where the first sum is over all nearest neighbors $\langle \bm r, \bm r' \rangle$ in $\mathcal A$ and $\bm S_{\bm r} = (\hat S_{\bm r}^x, \hat S_{\bm r}^y, \hat S_{\bm r}^z)$ are the spin-$1/2$ operators.
The numbers $h_{\bm r}$ are drawn randomly from the uniform probability distribution across $[-W, W]$ where $W$ is the disorder strength.
We ensure the rainbow scar is an eigenstate of Eq.\ \eqref{eq:general-Hamiltonian} by choosing the second term according to
\begin{equation}
	\hat H_{\mathcal B} = - \hat {\mathcal M} \hat H_{\mathcal A} \hat {\mathcal M}.
	\label{eq:H_B}
\end{equation}
Even though both $\hat H_{\mathcal A}$ and $\hat H_{\mathcal B}$ are disordered, Eq.\ \eqref{eq:scar-property} ensures that their sum annihilates the rainbow scar.
Note, however, that Eq.\ \eqref{eq:H_B} implies the random numbers $h_{\bm r}$ in parts $\mathcal A$ and $\mathcal B$ are not independent.
The operator $\hat H_{\mathcal A \mathcal B}$ connects parts $\mathcal A$ and $\mathcal B$.
We choose this operator such that the rainbow scar remains an exact eigenstate
\begin{equation}
	\hat H_{\mathcal A \mathcal B} = c \sum_{\substack{\bm r \in \mathcal A, \bm r' \in \mathcal B \\ \langle \bm r, \bm r' \rangle }} \bm S_{\bm r} \cdot \bm S_{\bm r'}.
	\label{eq:H_AB}
\end{equation}
The three operators $\hat H_{\mathcal A}$, $\hat H_{\mathcal B}$ and $\hat H_{\mathcal A \mathcal B}$ conserve the total magnetization in the $z$-direction.
Hence, the projection of the rainbow scar into a magnetization sector is itself an exact eigenstate.
These projections are degenerate with respect to $\hat H_{\mathcal A} \otimes \hat{\mathds{1}} + \hat{\mathds{1}} \otimes \hat H_{\mathcal B} + \hat H_{\mathcal A \mathcal B}$.
We lift the degeneracy by choosing the spectrum-generating term according to
\begin{align}
	\hat H_\text{SG} = \mu \sum_{\bm r \in \mathcal A \cup \mathcal B} \hat S_{\bm r}^z.
	\label{eq:H_SG}
\end{align}
The energy of $|\psi_\text{RB}^M \rangle$ is given by
\begin{equation}
	\hat H | \psi_\text{RB}^M \rangle = \left(\frac{1}{4} c L_y + \mu M\right) | \psi_\text{RB}^M \rangle.
\end{equation}
The states $\{|\psi_\text{RB}^M\rangle \}$ hence represent a tower of scar states with equal energy spacing.
Notice, that while the rainbow scar $|\psi_\text{RB}\rangle$ is an eigenstate of the first three terms in Eq.\ \eqref{eq:general-Hamiltonian}, it is not an eigenstate of $\hat{H}_\text{SG}$.

Figure~\ref{fig:entropy-vs-disorder} shows the entanglement entropy of eigenstates of $\hat H$ as a function of energy for several magnetization sectors.
We consider parameter values $J = \mu = c = 1$ and respectively weak and strong disorder.
At weak disorder, the entropy forms a narrow arc within each symmetry sector.
The scar states, on the other hand, generally appear as outliers external to each arc.
This behavior signals that the scar states are nonthermal.
When increasing the disorder strength, we observe the entropy of generic eigenstates decrease.
Note, that the entropy of the scar states is constant as a function of disorder strength.
Consequently, the entropy of the scar states is larger than that of generic eigenstates within the same symmetry sector at strong disorder.
We also observe other eigenstates with similar energy to the scar states and entropy at intermediate values.
We describe the origin of these high entropy eigenstates and investigate the behavior of the remaining spectrum at strong disorder in Sec.~\ref{sec:many-body-localization}.

Throughout this work, we generally consider parameter values $J = \mu = c = 1$ and the largest symmetry sector $M = 0$ unless otherwise stated.
We generally reach similar results for other values $J, \mu, c \neq 0$ and symmetry sectors.

\section{Localization}\label{sec:many-body-localization}
It is not immediately obvious how the model behaves with increasing disorder strength for a fixed system size.
On one hand, there exist numerous examples of spin models becoming MBL in a strongly disordered magnetic field \cite{Znidaric2008, Huse2010}.
In particular, a one-dimensional model similar to Eq.\ \eqref{eq:general-Hamiltonian} was shown to display MBL characteristics at strong disorder \cite{Srivatsa2022}.
On the other hand, the random magnetic fields are correlated which may represent an obstacle for localization.
Furthermore, MBL is believed to be unstable for two-dimensional systems in the thermodynamic limit \cite{Roeck2017Apr, Roeck2017Oct}.
In this section, we demonstrate that the model localizes at strong disorder and that the model has characteristics similar to localized models with scar states in one dimension.
Before tackling the general problem of characterizing the localization, we introduce the relevant concepts through a simple example.

\subsection{Simple example}\label{subsec:introducing-example}
It is convenient to rewrite the Hamiltonian to highlight the correlations in the magnetic fields $h_{\bm r}$
\begin{equation}
	\hat H = \hat H_0 + \sum_{\bm r \in \mathcal A} h_{\bm r} \hat D_{\bm r}
	\label{eq:disorder-highlighted-Hamiltonian}
\end{equation}
where we denote $\hat D_{\bm r} = \hat S_{\bm r}^z - \hat S_{\tilde{\bm r}}^z$ as ``disorder operators'' and $\hat H_0$ is given by
\begin{equation}
	\hat H_0 = J \bigg(
			\sum_{\substack{\bm r, \bm r' \in \mathcal A \\ \langle \bm r, \bm r' \rangle}} \bm S_{\bm r} \cdot \bm S_{\bm r'}
			- \sum_{\substack{\bm r, \bm r' \in \mathcal B \\ \langle \bm r, \bm r' \rangle}} \bm S_{\bm r} \cdot \bm S_{\bm r'} 
		\bigg)
		+ \hat H_{\mathcal A \mathcal B}
		+ \hat H_\text{SG}.
\end{equation}

In MBL systems, the set of disorder operators is commonly chosen to act uniquely on each state in a basis, i.e., for a set of disorder operators $\{\hat D_i \}$ and a basis $\{| \varphi \rangle \}$, if $\langle \varphi | \hat D_i | \varphi \rangle = \langle \varphi' | \hat D_i | \varphi' \rangle$ for all $i$, then $| \varphi \rangle = | \varphi' \rangle$.
This is not true for our model in Eq.~\eqref{eq:disorder-highlighted-Hamiltonian}.
Here, the disorder operators have the same action on some basis states.
For instance, consider a lattice of size $4 \times 2$.
For any disorder realization $\{h_{\bm r} | \bm r \in \mathcal A\}$, we have
\begin{equation}
\begin{split}
	\bigg\langle 
	\begin{matrix}
		\downarrow & \downarrow & \downarrow & \uparrow \\
		\uparrow & \uparrow & \downarrow & \uparrow
	\end{matrix}
	\bigg|
	\sum_{\bm r \in \mathcal A} h_{\bm r} \hat D_{\bm r}
	\bigg|
	\begin{matrix}
		\downarrow & \downarrow & \downarrow & \uparrow \\
		\uparrow & \uparrow & \downarrow & \uparrow
	\end{matrix}
	\bigg\rangle
	= \\
	\bigg \langle 
	\begin{matrix}
		\downarrow & \uparrow & \uparrow & \uparrow \\
		\downarrow & \uparrow & \downarrow & \downarrow
	\end{matrix}
	\bigg |
	\sum_{\bm r \in \mathcal A} h_{\bm r} \hat D_{\bm r}
	\bigg |
	\begin{matrix}
		\downarrow & \uparrow & \uparrow & \uparrow \\
		\downarrow & \uparrow & \downarrow & \downarrow
	\end{matrix}
	\bigg \rangle.
\end{split}
\label{eq:introducing-example-action}
\end{equation}
Hence, the disorder operators $\{\hat D_{\bm r}\}$ do not distinguish between the two basis states.
Intuitively, this behavior is caused by pairs of spins being mirror symmetric.
In the above example, sites $\bm r_1 = (0, 0)$, $\tilde {\bm r}_1 = (3, 0)$ and $\bm r_2 = (1, 1)$, $\tilde {\bm r}_2 = (2, 1)$ are mirror symmetric.
The disorder operators act trivially on these pairs of spins 
\begin{equation}
	(h_{\bm r_1} \hat D_{\bm r_1} + h_{\bm r_2} \hat D_{\bm r_2})	
	\bigg|
	\begin{matrix}
		\downarrow & \downarrow & \downarrow & \uparrow \\
		\uparrow & \uparrow & \downarrow & \uparrow
	\end{matrix}
	\bigg\rangle
	= 0.
\end{equation}
The same is true for the corresponding basis state with both pairs flipped.
The fact that the disorder operators do not distinguish between two basis states manifests itself in the structure of the energy eigenstates at strong disorder.
We explore this structure in Sec.\ \ref{subsec:general-characterization}.

In the above example, the disorder operators had the same action on two different basis states within the zero magnetization sector.
The discussion naturally extends to the case where more than two basis states are treated identically by the disorder operators.
In this case, more than two pairs of spins are mirror symmetric and flipping any two pairs yields a new basis state which can not be distinguished by the disorder operators.

\subsection{General characterization of localization}\label{subsec:general-characterization}
Let $| \bm D, n \rangle$ be a simultaneous eigenket of the disorder operators $\hat D_{\bm r}$ where $\bm D = \{D_{\bm r} | \bm r \in \mathcal A\}$ are the corresponding eigenvalues, i.e., $\hat D_{\bm r} | \bm D, n \rangle = D_{\bm r} | \bm D, n \rangle$ for all $\bm r \in \mathcal A$.
We refer to $\bm D$ as the ``disorder indices''.
The disorder indices do not fully describe a state and the extra index $n$ distinguishes between states with identical $\bm D$.

At strong disorder, the last term in Eq.\ \eqref{eq:disorder-highlighted-Hamiltonian} dominates and we expect each energy eigenstate to have significant support only on basis states with identical disorder indices.
We refer to this behavior as eigenstates localizing within a subspace $\mathcal V_{\bm D} = \mathrm{span}(\{| \bm D, n \rangle| n = 1, 2, \ldots \})$.
Note, however, that this behavior may not persist at any finite disorder strength in the thermodynamic limit.
The energy eigenstates are labeled by $\bm D$ at strong disorder.
We further introduce an index $m$ to distinguish energy eigenstates localizing within the same subspace $\mathcal V_{\bm D}$
\begin{equation}
	| E_{\bm D, m} \rangle \approx \sum_n c_{\bm D; mn}|\bm D, n \rangle.
	\label{eq:eigenstate-expansion}
\end{equation}
This expression is an approximation rather than an equality due to the small support on other subspaces $\mathcal V_{\bm D'}$ with $\bm D' \neq \bm D$ which decreases with increasing disorder strength.

The special case where the disorder operators act uniquely on a basis state corresponds to $\dim(\mathcal V_{\bm D}) = 1$.
At large disorder, the corresponding eigenstate has significant support on a single basis state which we refer to as strong localization.

\begin{figure}
	\includegraphics[width=\linewidth]{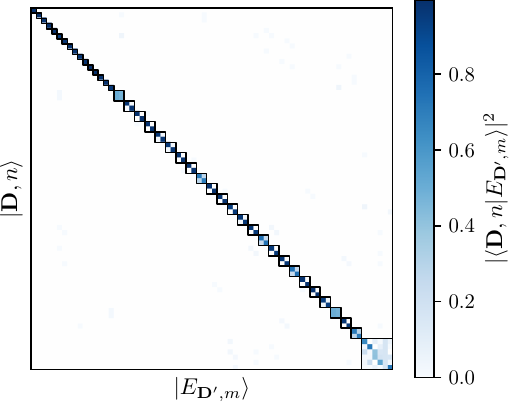}
\caption{
	The energy eigenstates of a single disorder realization for system size $4 \times 2$, disorder strength $W = 14$, parameters $J = c = \mu = 1$ and in the magnetization sector $M = 0$.
	Each row corresponds to a basis state $|\bm D, n \rangle$ and each column to an eigenstate $|E_{\bm D', m} \rangle$.
	The color intensity of each pixel displays the norm squared overlap between the corresponding basis state and eigenstate $|\langle \bm D, n | E_{\bm D', m} \rangle |^2$.
	The eigenstates are rearranged to allow the diagonal shape and hence are not sorted into ascending order with respect to energy.
	The square boxes show the subspaces $\mathcal V_{\bm D}$ where the eigenstates are expected to localize according to Eq.~\eqref{eq:eigenstate-expansion}.
	The figure illustrates that most eigenstates localize on smaller subspaces within each $\mathcal V_{\bm D}$.
	Similar results are found for larger system sizes.
}
	\label{fig:localization}
\end{figure}

In the case $\mathrm{dim}(\mathcal V_{\bm D}) > 1$, the structure of the energy eigenstates is not immediately obvious.
One might fear that all coefficients in Eq.~\eqref{eq:eigenstate-expansion} are nonzero $c_{\bm D; mn} \neq 0$ and the system ``partially localizes'' \cite{Iversen2023}.
It turns out, however, that the majority of energy eigenstates localize more strongly than predicted by Eq.~\eqref{eq:eigenstate-expansion}.
Figure \ref{fig:localization} illustrates the localization for system size $4 \times 2$.
The figure demonstrates that eigenstates generally localize on a smaller subspace within each $\mathcal V_{\bm D}$.
For instance, eigenstates in subspaces with $\dim(\mathcal V_{\bm D}) = 2$ tend to localize on a single basis state.
We find similar results for larger system sizes.

We understand the strong localization of eigenstates from degenerate perturbation theory.
The central observation is that the offdiagonal matrix elements $\langle \bm D, n | \hat H_0^p | \bm D, n' \rangle$ generally vanish for small powers $p$.
In other words, the basis states with identical disorder indices are not connected directly by $\hat H_0^p$.
Consequently, these states do not tend to mix and the energy eigenstates have significant support on only a few basis states.
We present a detailed analysis in Appendix~\ref{sec:appendix}.

While the eigenstates localize more strongly than predicted by Eq.~\eqref{eq:eigenstate-expansion}, they might not localize on a single basis state.
However, the energy difference between eigenstates with significant support within the same subspace $\mathcal V_{\bm D}$ vanishes with increasing disorder strength.
Consequently, the dynamics arising from such eigenstates can be made arbitrarily slow for sufficiently strong disorder.

Let $\bm D_0$ be the disorder indices given by $[\bm D_0]_{\bm r} = 0$ for all $\bm r \in \mathcal A$.
The perturbative arguments are valid for all subspaces except $\mathcal V_{\bm D_0}$.
This subspace contains the projections of the rainbow scar as well as other eigenstates.
By construction, this subspace evades disorder and the entropy of eigenstates in $\mathcal V_{\bm D_0}$ is typically larger than that of localized states but smaller or equal to the entropy of the scar states.
The subspace $\mathcal V_{\bm D_0}$ hence represents a nonlocalized subspace embedded in an otherwise localized spectrum.
The subspace is visible in Fig.~\ref{fig:entropy-vs-disorder}(b) as the eigenstates with energy close to a scar state and intermediate values of entropy.
The part of ${\mathcal V}_{\bm D_0}$ with magnetization $M = 0$ is also visible in the bottom right corner of Fig.~\ref{fig:localization}.
In the following, we include the symmetry sector in the notation, i.e., the subspace with disorder indices $\bm D$ and magnetization $M$ is denoted as $\mathcal V_{\bm D}^M$.
We determine the dimension of the nonlocalized subspace across all symmetry sectors $\cup_M \mathcal V_{\bm D_0}^M$ by noting that it consists of all mirror symmetric basis states.
These basis states are fully determined by the spins in part $\mathcal A$ and the dimension is given by $\mathrm{dim}(\cup_M \mathcal V_{\bm D_0}^M) = 2^{L_x L_y / 2}$.
Consistent with Ref.~\cite{Srivatsa2022}, we find that the nonlocalized eigenstates represent a vanishing small part of the full Hilbert space $\lim_{L_x, L_y \to \infty}\big[\mathrm{dim}(\cup_M \mathcal V_{\bm D_0}^M) / \mathrm{dim}(\mathcal H)\big] = 0$.

The considerations presented in this section are not specific to our model.
They can be taken as general guidelines for constructing strongly localized models hosting QMBS.
Strong localization is obtained by ensuring the nondisordered part of the Hamiltonian restricted to the subspace $\mathcal V_{\bm D}$ is diagonal.
This rule may help future work on disordered models hosting QMBS achieve stronger localization. 

\subsection{Adjacent gap ratio}\label{subsec:adjacent-gap-ratio}
\begin{figure*}
	\includegraphics[width=\linewidth]{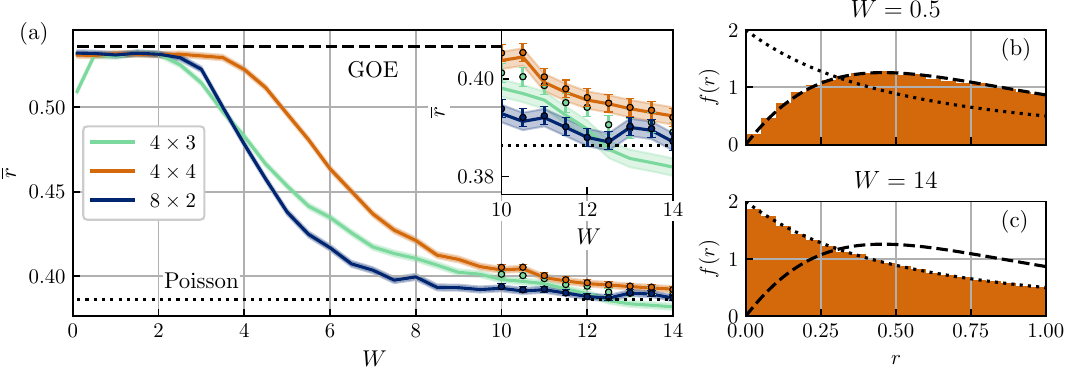}
	\caption{
		(a) The adjacent gap ratio $\overline r$ averaged over the $10^2$ energies closest to $(E_\text{min} + E_\text{max}) / 2$ in $10^3$ disorder realizations (solid lines) as a function of disorder strength $W$ for different system sizes $L_x \times L_y$.
		The GOE and Poisson values are illustrated as dashed and dotted horizontal lines.
		The shaded areas display two standard deviations on the estimate of the mean when assuming a Gaussian distribution.
		At weak disorder, $\overline r$ coincides with GOE indicating that the system is thermal.
		At strong disorder, $\overline r$ agrees with the Poisson distribution showing that the system is localized.
		The adjacent gap ratio decreases below the Poisson value for system size $4 \times 3$ at strong disorder $12 \lesssim W$ as illustrated in the inset.
		When restricting the computation of $\overline r$ to eigenstates with different disorder indices (dots), the adjacent gap ratio converges to the Poisson value for all system sizes.
		The error bars display two standard deviations on the estimate of the mean.
		(b)-(c) The distribution of adjacent gap ratio for system size $4 \times 4$ at (b) weak disorder $W = 0.5$ and (c) strong disorder $W = 14$.
		The distributions $f_\text{GOE}$ and $f_\text{Poisson}$ from Eq.\ \eqref{eq:probability-density-functions} are illustrated as dashed and dotted curves.
		The data agrees with (b) $f_\text{GOE}$ at weak disorder showing that the system is thermal and agrees with (c) $f_\text{Poisson}$ at strong disorder indicating that the system is localized.
		In all panels, we consider parameters $J = c = \mu = 1$ and the symmetry sector $M = 0$.
		}\label{fig:adjacent-gap-ratio}
\end{figure*}
The distribution of energy levels indicates whether the system is thermal or localized.
The distribution follows the Wigner surmise in the thermal phase.
In particular, the distribution is given by the Gaussian orthogonal ensemble (GOE) since the Hamiltonian in Eq.\ \eqref{eq:general-Hamiltonian} is invariant under time reversal.
We expect the energy levels to follow the Poisson distribution at strong disorder similar to an MBL system \cite{Serbyn2016}.
Let $\{E_i\}$ be the energies in ascending order and let $\delta_i = E_{i + 1} - E_i > 0$ be the $i$-th energy gap.
We consider the adjacent gap ratio \cite{Huse2007}
\begin{equation}
	r_i = \frac{\min(\delta_i, \delta_{i + 1})}{\max(\delta_i, \delta_{i + 1})}.
\end{equation}
The transition from thermal to localized behavior is identified by studying the distribution of the adjacent gap ratio.
The probability density functions of $r$ for a thermal and localized system are given by \cite{Atas2013}
\begin{subequations}
\begin{gather}
	f_\text{GOE}(r) = \frac{27}{4} \frac{r(1 + r)}{(1 + r + r^2) ^ {5 / 2}}, \\
	f_\text{Poisson}(r) = \frac{2}{(1 + r) ^ 2}.
\end{gather}\label{eq:probability-density-functions}%
\end{subequations}
The corresponding expectation values of the two distributions are given by $\langle r \rangle_\text{GOE} = 2(2 - \sqrt 3) \approx 0.536$ and $\langle r \rangle_\text{Poisson} = 2\ln 2 - 1 \approx 0.386$.

We consider $10^3$ disorder realizations and compute the adjacent gap ratio from the $10^2$ energies closest to the center of the spectrum, i.e., $(E_\text{min} + E_\text{max}) / 2$ where $E_\text{min}$ and $E_\text{max}$ are respectively the smallest and largest energy in the spectrum.
The adjacent gap ratio is averaged over both disorder realizations and this central part of the spectrum.
Figure~\ref{fig:adjacent-gap-ratio}(a) illustrates the mean adjacent gap ratio $\overline r$ as a function of disorder strength $W$ for different system sizes $L_x \times L_y$.
At weak disorder $0.1 \lesssim W \lesssim 2$, the mean adjacent gap ratio agrees with GOE for all system sizes.
Hence, the system is thermal at weak disorder.
When increasing the disorder strength $2 \lesssim W \lesssim 12$, $\overline r$ departs from the GOE value and approaches the Poisson value.
At strong disorder $12 \lesssim W$, the mean adjacent gap ratio agrees well with the Poisson value indicating that the model localizes.

For system size $4 \times 3$ at strong disorder $12 \lesssim W$, the adjacent gap ratio is seen to decrease below the Poisson value.
This behavior is explained by the model not localizing analogously to conventional MBL.
As discussed in Sec.~\ref{subsec:general-characterization}, the energy gap between eigenstates with identical disorder indices decreases with increasing disorder strength.
Therefore, the distribution of energy levels only approaches the Poisson distribution when considering eigenstates with different disorder indices. 
With this in mind, we analyze the adjacent gap ratio more carefully by only including energy gaps $\delta_i$ of eigenstates with different disorder indices, i.e., pairs of adjacent eigenstates $|E_{\bm D, m} \rangle$ and $|E_{\bm D', m'}\rangle$ with $\bm D \neq \bm D'$.
This filtering is only possible at strong disorder where all eigenstates are sufficiently localized and the disorder indices of each eigenstate can be reliably determined numerically.
For each eigenstate $|E_i\rangle$, we determine the basis states $|\bm D, n \rangle$ with the largest norm squared overlap and thereby deduce the disorder indices, i.e., $\mathrm{argmax}_{\bm D}(\sum_n |\langle \bm D, n | E_i \rangle|^2)$.
As illustrated in Fig.~\ref{fig:adjacent-gap-ratio}(a), the adjacent gap ratio of all system sizes converges to the Poisson value when only considering eigenstates of different disorder indices.
The two methods of calculating the adjacent gap ratio nearly coincide for system sizes $4 \times 4$ and $8 \times 2$.

Finally, we illustrate the adjacent gap ratio distribution in Figs.~\ref{fig:adjacent-gap-ratio}(b)-(c) and compare it with Eq.\ \eqref{eq:probability-density-functions}.
The distribution agrees with $f_\text{GOE}$ at weak disorder and with $f_\text{Poisson}$ at strong disorder.
Thus, the system transitions from the thermal phase to being localized as disorder is introduced.

\subsection{Entanglement entropy}\label{subsec:entanglement-entropy}
\begin{figure}
	\includegraphics{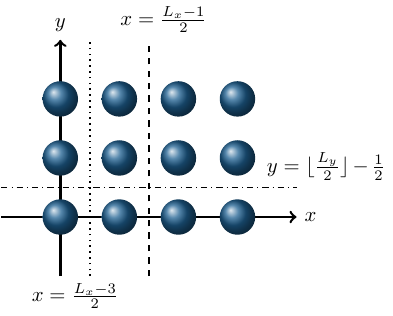}
	\caption{
	The system is separated into two parts $A$ and $B$.
	We illustrate the separation for system size $4 \times 3$.
	The boundary of the two parts is described by the lines $x = (L_x - 1) / 2$ (dashed line), $x = (L_x - 3) / 2$ (dotted line) and $y = \lfloor L_y / 2 \rfloor - 1 / 2$ (dash-dotted line).
	}\label{fig:partitions}
\end{figure}
\begin{figure}
	\includegraphics[width=\linewidth]{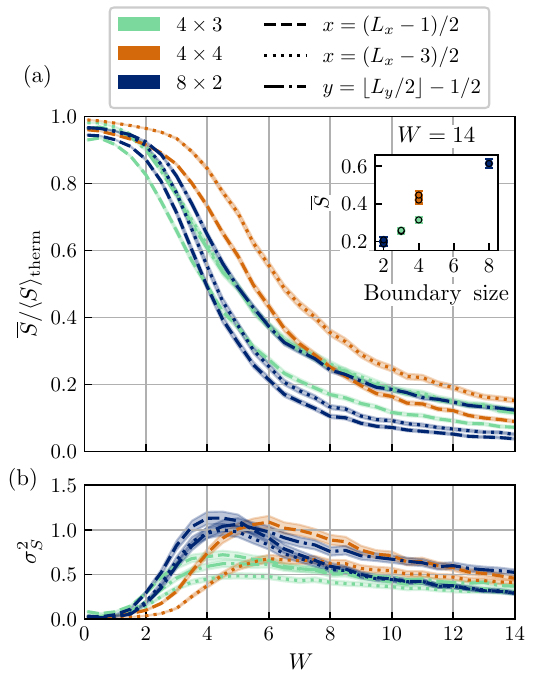}
	\caption{
		(a) The average entanglement entropy $\overline S$ over $2\times 10^3$ disorder realizations of the eigenstate closest in energy to $(E_\text{min} + E_\text{max}) / 2$.
		The entropy is plotted as a function of disorder strength $W$ for different system sizes and partitions.
		The system size is indicated by color and the partition by line style.
		At weak disorder, the entropy for all considered system sizes and partitions agree with $\langle S \rangle_\text{therm}$.
		At strong disorder, the entropy departs from the thermal value and instead displays area-law scaling as illustrated in the inset.
		(b) The sample variance as a function of disorder strength.
		The variance displays a peak at intermediate disorder strength indicating that the system transitions from being thermal to localized.
		The shaded areas and error bars in both figures display two standard deviations on the estimate of the mean and variance. 
		We consider parameters $J = c = \mu = 1$ and the symmetry sector $M = 0$.
		}\label{fig:entanglement-entropy}
\end{figure}
We further establish the transition from the thermal phase to localization by studying the von Neumann entropy.
The system is separated into two parts $A$ and $B$.
Let $\rho$ be the density matrix describing the full system and let $\rho_A = \mathrm{Tr}_B(\rho)$ be the reduced density matrix of part $A$.
The von Neumann entropy is given by
\begin{equation}
	S = - \mathrm{Tr}[\rho_A \ln(\rho_A)].
\end{equation}
In the thermal phase, the entropy scales with the volume of the system.
In particular, for a thermal system with tensor product structure $\mathcal H = \mathcal H_A \otimes \mathcal H_B$, the entropy of an infinite temperature eigenstate is given by the Page value $\langle S \rangle_\text{Page}$ \cite{Page1993}.
The model described by Eqs.\ \eqref{eq:general-Hamiltonian}-\eqref{eq:H_SG} conserves the total magnetization and the Hilbert space of a generic magnetization sector $\mathcal H_M$ does not have tensor product structure $\mathcal H_M \neq \mathcal H_A \otimes \mathcal H_B$.
Therefore, the Page value is not representative of the average entropy of a thermal state in this model. 
Instead, the expected entropy in the thermal phase is given by \cite{Bianchi2019, Bianchi2022}
\begin{equation}
\begin{split}
	\langle S \rangle_\text{therm}
	=&\sum_{M_A = \max\big(- \frac{N_A}{2}, M - \frac{N_B}{2}\big)}^{\min\big(\frac{N_A}{2}, M + \frac{N_B}{2}\big)} \frac{\dim(\mathcal H_{M_A}) \dim(\mathcal H_{M_B})}{\dim(\mathcal H_M)} \\
	&\times \big\{\langle S \rangle_\text{Page}
	+ \Psi[\dim(\mathcal H_M) + 1] \\
	&- \Psi[\dim(\mathcal H_{M_A}) \dim(\mathcal H_{M_B}) + 1]\big\}.
\end{split}
\label{eq:entropy-thermal}
\end{equation}
In this expression, $M_A$ is the magnetization of part $A$ and $M_B = M - M_A$ is the magnetization of part $B$.
The Hilbert space of part $A$ ($B$) with magnetization $M_A$ ($M_B$) is denoted by $\mathcal H_{M_A}$ ($\mathcal H_{M_B}$).
The number of lattice sites in part $A$ ($B$) is denoted by $N_A$ ($N_B$).
Finally, $\Psi$ is the digamma function.
In the MBL phase, on the other hand, the entanglement entropy is proportional to the boundary of the partition \cite{Abanin2013, Bauer2013}.
Furthermore, the variance of entropy displays a peak at the transition between the thermal and MBL phase \cite{Kjall2014, Luitz2015}.
We expect our model to display similar characteristics of entanglement entropy at strong disorder.

We consider $2\times 10^3$ disorder realizations and for each realization compute the entropy of the eigenstate closest in energy to $(E_\text{min} + E_\text{max}) / 2$.
The sample average $\overline S$ and sample variance $\sigma_S^2$ are then determined.
Figure \ref{fig:entanglement-entropy}(a) illustrates the average entropy as a function of disorder strength for different system sizes and partitions $A, B$.
We consider three partitions described by the lines $x = (L_x - 1) / 2$, $x = (L_x - 3) / 2$ and $y = \lfloor L_y / 2 \rfloor - 1 / 2$ where $\lfloor \cdot \rfloor$ is the function that rounds down to the nearest integer.
All sites on one side of the line represent part $A$ and the remaining sites represent part $B$.
The partitions are illustrated in Fig.\ \ref{fig:partitions} for system size $4 \times 3$.

Figure~\ref{fig:entanglement-entropy}(a) illustrates the average entropy as a function of disorder strength and Fig.~\ref{fig:entanglement-entropy}(b) shows the variance.
In both figures, we display multiple system sizes and partitions.
For weak disorder, the average entropy agrees with $\langle S \rangle_\text{therm}$ for all system sizes and partitions.
As the disorder strength is increased, the average entropy decreases and the variance displays a sharp peak.
The large variance signals that the system transitions from the thermal phase to being localized.\@
In the inset of Fig.~\ref{fig:entanglement-entropy}(a), we display the entropy as a function of boundary size for each system size and partition at $W = 14$.
We observe that the entropy is proportional to the size of the boundary as expected for localization.

\section{Scar dynamics}\label{sec:scar-dynamics}
The presence of rainbow scars in the model causes nonthermal dynamics.
We observe such dynamics by choosing the initial state with large support in the scar subspace.
We consider the thermofield double state $|\psi_\text{TFD} (\beta)\rangle$ at inverse temperature $\beta$ \cite{Cottrell2019}.
Let $\hat H_\text{TFD}$ be an operator acting on $\mathcal{H}_\mathcal{A}$, then the thermofield double state is given by
\begin{equation}
	| \psi_\text{TFD}(\beta) \rangle = \frac{1}{\sqrt{Z}} \sum_i e^{-\beta E_i / 2} |E_i \rangle \otimes \hat{\mathcal{M}}|E_i \rangle
\end{equation}
where $E_i$ and $|E_i\rangle$ are the energies and eigenstates of $\hat H_\text{TFD}$ and $Z$ is a normalization constant.
Several protocols have been proposed for realizing $|\psi_\text{TFD}(\beta)\rangle$ \cite{Cottrell2019, Wu2019, Zhu2020}.
The thermofield double state reduces to the rainbow scar from Eq.~\eqref{eq:general-rainbow-scar} in the limit of infinite temperature $|\psi_\text{TFD}(0)\rangle = |\psi_\text{RB}\rangle$ for any operator ${\hat H}_\text{TFD}$.

\begin{figure}
	\includegraphics[width=\linewidth]{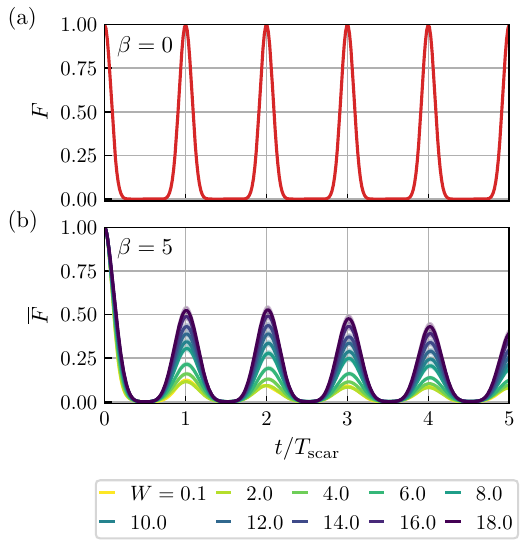}
	\caption{
		(a) The fidelity $F$ as a function of time $t$ when initializing the system as the rainbow scar $|\psi(0) \rangle = |\psi_\text{RB}\rangle$.
		The fidelity displays revivals because the projections $|\psi_\text{RB}^M\rangle$ have equal energy spacing.
		(b) The system is initialized as the thermofield double state $|\psi(0)\rangle = |\psi_\text{TFD}(\beta)\rangle$ with $\beta = 5$ and time evolved at different disorder strengths.
		The figure displays the average fidelity over $500$ disorder realizations and the shaded areas show two standard deviations on the estimate of the mean.
		At weak disorder, the fidelity displays revivals with a small amplitude.
		The revival amplitude increases with increasing disorder strength.
		In both panels, we consider system size $L_x \times L_y = 4 \times 3$ and parameters $J = c = \mu = 1$.
	}\label{fig:scar-dynamics}
\end{figure}

First, we consider the initial state $|\psi(t=0)\rangle = |\psi_\text{RB}\rangle$ and study the system dynamics.
The state is time-evolved according to $|\psi(t) \rangle = e^{-i\hat H t}|\psi(0)\rangle$ and the fidelity $F(t) = |\langle \psi(0) | \psi(t) \rangle |^2$ is determined.
Figure~\ref{fig:scar-dynamics}(a) shows the fidelity as a function of time.
As expected, $F(t)$ displays revivals with period $T_\text{scar} = \pi/\mu$ since the initial state resides fully in the scar subspace.

Next, we initialize the system only partially within the scar subspace $|\psi(0) \rangle = |\psi_\text{TFD}(\beta)\rangle$ with $\beta = 5$.
The thermofield double state is prepared with respect to the operator
\begin{equation}
	\hat H_\text{TFD} = J \sum_{\substack{\bm r, \bm r' \in \mathcal A \\ \langle \bm r, \bm r' \rangle}} \hat S_{\bm r}^z\hat S_{\bm r'}^z + h \sum_{\bm r \in \mathcal A} \hat S_{\bm r}^z
\end{equation}
with $J = h = 1$.
Since $|\psi_\text{TFD}(\beta)\rangle$ only partially resides in the scar subspace, we do not expect perfect revivals.
Figure~\ref{fig:scar-dynamics}(b) illustrates the average fidelity over $500$ disorder realizations for different disorder strengths.
At weak disorder, the fidelity displays revivals with a small amplitude.
The reduced amplitude is caused by the initial state having significant support outside the scar subspace.
At strong disorder, the fidelity displays revivals with a larger amplitude than the thermal case.
This behavior is explained by the initial state having support on basis states $|\bm D_0, n\rangle$ with identical disorder indices to the rainbow scar.
Consequently, the initial state is a linear combination of energy eigenstates which all have energy close to a scar state.
Thus, the disorder enhances the scar revivals from initial states with partial support in the scar subspace.

\section{Alternative model}\label{sec:tower-of-rotated-EPR-scars}
Inverted quantum scars are characterized by having a larger entanglement entropy than the background of localized eigenstates.
The entanglement entropy of the scar states from Eq.~\eqref{eq:projections-rainbow-scar} displays logarithmic scaling with system size for a particular bipartition. 
The entropy of the scar states for this bipartition may therefore be similar to that of localized states for small system sizes.
We address this issue by constructing a different tower of scar states with larger entanglement entropy for generic bipartitions.
The model follows the same theoretical framework presented in Appendix~\ref{appendix:scar-state-framework} and hence also illustrates the flexibility in our construction.

\subsection{Projections of the scar state}\label{subsec:projections-of-the-rotated-EPR-scar}
\begin{figure}
	\centering
	\includegraphics{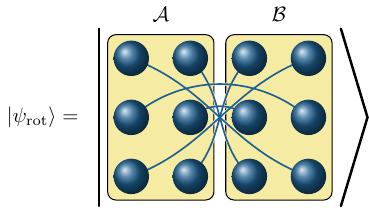}
	\caption{
		Illustration of the state from Eq.~\eqref{eq:rotated-scar-bell-states} for a lattice of size $4 \times 3$.
		The state is the tensor product of Bell states $(|\mathord \downarrow \rangle_{\bm r} |\mathord \downarrow \rangle_{\bar{\bm r}} + |\mathord \uparrow \rangle_{\bm r} |\mathord\uparrow \rangle_{\bar{\bm r}})/\sqrt{2}$ where $\bar{\bm r}$ corresponds to rotating $\bm r$ around the center of the lattice $\bm{r}_\text{center}$ by an angle $\pi$.
		Two balls connected by a line represent a Bell state.
		}
	\label{fig:rotated-scar-illustration}
\end{figure}

We construct a scar state which, similar to the rainbow scar, is the tensor product of Bell pairs when the system consists of spin-$1/2$ particles.
We ensure this scar state has larger entanglement entropy for generic bipartitions than the rainbow scar by increasing the distance between the two sites constituting each Bell pair.
Consider the operator $\hat{\mathcal R}_{\pi}$ which rotates all lattice sites around the center of the lattice ${\bm r}_\text{center} = [(L_x - 1)/ 2, (L_y - 1) / 2]$ by an angle $\pi$ without flipping any spins.
We consider the state
\begin{equation}
	|\psi_\text{rot}^\text{general} \rangle = \frac{1}{\sqrt{\dim(\mathcal H_{\mathcal A})}}\sum_{|\varphi \rangle \in \mathrm{basis}(\mathcal H_{\mathcal A})} | \varphi \rangle \otimes \hat {\mathcal R}_{\pi} | \varphi \rangle.
	\label{eq:rotated-scar}
\end{equation}
Similar to the rainbow scar, this state has a special property.
For any operator $\hat{\mathcal O}$ acting on $\mathcal{H}_{\mathcal A}$ we have
\begin{equation}
	(\hat{\mathcal O} \otimes \hat{\mathds 1})| \psi_\text{rot}^\text{general} \rangle 
	= (\hat{\mathds 1} \otimes \hat {\mathcal R}_{\pi} \hat{\mathcal O}^T \hat {\mathcal R}_{\pi})|\psi_\text{rot}^\text{general} \rangle.
	\label{eq:general-scar-property}
\end{equation}
When the system consists of spin-$1/2$ particles, the state may be written as a tensor product of Bell states 
\begin{equation}
	|\psi_\text{rot} \rangle = \bigotimes_{\bm r \in \mathcal A} \left( \frac{|\mathord \downarrow \rangle_{\bm r} \otimes |\mathord \downarrow \rangle_{\bar{\bm r}} + |\mathord \uparrow \rangle_{\bm r} \otimes |\mathord\uparrow \rangle_{\bar{\bm r}}}{\sqrt 2} \right)
	\label{eq:rotated-scar-bell-states}
\end{equation}
with $\bm{r} = (x, y)$ and $\bar{\bm{r}} = (L_x - 1 - x, L_y - 1 - y)$ is obtained by rotating $\bm{r}$ around the center of the lattice ${\bm r}_\text{center}$ by an angle $\pi$.
Figure~\ref{fig:rotated-scar-illustration} illustrates the state from Eq.~\eqref{eq:rotated-scar-bell-states}.

Consider the normalized projection of Eq.~\eqref{eq:rotated-scar-bell-states} into the subspace with magnetization $M$ in the $z$-direction
\begin{equation}
|\psi_\text{rot}^M\rangle = \frac{\hat{P}_M |\psi_\text{rot}\rangle}{\sqrt{\langle \psi_\text{rot} | \hat P_M | \psi_\text{rot} \rangle}}.
\end{equation}
We focus on the scar states 
\begin{equation}\label{eq:alternative-scar-states}
	\Big\{|\psi_\text{rot}^M\rangle \Big| M = -\frac{L_x L_y}{2}, -\frac{L_x L_y}{2} + 2, \ldots, \frac{L_x L_y}{2}\Big\}
\end{equation}
since ${|\psi_\text{rot}^M\rangle = 0}$ for $M = - L_x L_y / 2 + 1, - L_x L_y / 2 + 3, \ldots, L_x L_y / 2 - 1$.

\subsection{Entanglement entropy of the scar states}
\label{subsec:entanglement-entropy-of-EPR-scar}
The entanglement entropy of the scar states from Eq.~\eqref{eq:alternative-scar-states} display volume-law scaling with system size for a variety of bipartitions.
In the following, we partition the system into two parts $A_\ell$ and $B_\ell$ using a line $\ell$.
For a given line $\ell$, all lattice sites on one side of the line belong to $A_\ell$ and the remaining sites belong to $B_\ell$.
We focus on lines not passing through any lattice points.
Consider the line $\ell_{{\bm r}_\text{dir}}: \{\bm{r}_\text{center} + s {\bm r}_\text{dir}| s \in \mathbb R\}$ which passes through the center of the lattice ${\bm r}_\text{center}$ in the direction ${\bm r}_\text{dir}$.
Notice that for any ${\bm r}_\text{dir}$, the line $\ell_{\bm{r}_\text{dir}}$ separates all Bell pairs.
Consequently, the entanglement entropy of $|\psi_\text{rot}\rangle$ is given by $S = L_x L_y \ln(2) / 2$ and hence displays volume-law scaling for any $\bm{r}_\text{dir}$.
Furthermore, the entanglement entropy of the projections $|\psi_\text{rot}^M\rangle$ also displays volume-law scaling for any such bipartition.

We compare the entanglement entropy of the scar states in Eq.~\eqref{eq:alternative-scar-states} with the scar states based on the rainbow scar in Eq.~\eqref{eq:projections-rainbow-scar}.
The distance between two lattice sites forming a Bell pair in $|\psi_\text{rot}\rangle$ is equal to or larger than the corresponding Bell pair in $|\psi_\text{RB}\rangle$.
Consequently, the probability that a randomly chosen bipartition separates a Bell pair is greater for $|\psi_\text{rot}\rangle$ than for $|\psi_\text{RB}\rangle$.
The entanglement entropy of Eq.~\eqref{eq:alternative-scar-states} is hence larger than the entropy of Eq.~\eqref{eq:projections-rainbow-scar} for a generic bipartition.
This means that the scar states $|\psi_\text{rot}^M\rangle$ are more easily distinguished from localized states by the entanglement entropy than $|\psi_\text{RB}^M\rangle$.

\subsection{Parent Hamiltonian}
\label{subsec:parent-Hamiltonian-of-EPR-scar}
We construct a parent Hamiltonian for the scar states $|\psi_\text{rot}^M\rangle$ by following the ideas presented in Sec.~\ref{subsec:parent-hamiltonian} and Appendix~\ref{appendix:scar-state-framework}.
Consider the Hamiltonian
\begin{align}
	\hat H_\text{rot} = \hat H_{\mathcal A} \otimes \hat {\mathds 1} + \hat {\mathds 1} \otimes \tilde H_{\mathcal B} + \tilde H_{\mathcal A \mathcal B} + \hat H_\text{SG}.
	\label{eq:parent-Hamiltonian-rotated-EPR-state}
\end{align}
where $\hat H_{\mathcal A}$ and $\hat H_\text{SG}$ are identical to Eqs.~\eqref{eq:HA} and \eqref{eq:H_SG}.
The remaining terms are given by
\begin{align}
	&\tilde H_{\mathcal B} = - \hat {\mathcal R}_{\pi} \hat H_{\mathcal A} \hat {\mathcal R}_{\pi},\\
	&\tilde H_{\mathcal A \mathcal B} = c \Big[ \sum_{\substack{\bm r \in \mathcal A, \bm r' \in \mathcal B \\ \langle \bm r, \bm r' \rangle \\ y, y' \leq \frac{L_y - 1}{2}}} \big(\bm S_{\bm r} \cdot \bm S_{\bm r'}\big) - \sum_{\substack{\bm r \in \mathcal A, \bm r' \in \mathcal B \\ \langle \bm r, \bm r' \rangle \\ \frac{L_y - 1}{2} < y, y'}} \big( \bm S_{\bm r} \cdot \bm S_{\bm r'}\big) \Big],
	\label{eq:H_AB^tilde}
\end{align}
where $y, y'$ are the $y$-coordinates of respectively $\bm{r}$ and $\bm{r}'$.
The first sum in Eq.~\eqref{eq:H_AB^tilde} includes all nearest neighbors across the boundary between $\mathcal A$ and $\mathcal B$ which lies in the lower part of the lattice.
Similarly, the second sum includes all nearest neighbors across the boundary in the upper part of the lattice.
The states $|\psi_\text{rot}^M\rangle$ are eigenstates of $\hat{H}_\text{rot}$ with energy depending on whether $L_y$ is even or odd
\begin{subequations}
\begin{align}
	\hat H_\text{rot} | \psi_\text{rot}^M \rangle &= \mu M | \psi_\text{rot}^M \rangle, \quad &&\text{($L_y$ even)}\label{eq:alternative-scar-state-energies-even}\\
	\hat H_\text{rot} | \psi_\text{rot}^M \rangle &= \Big(\frac{c}{4} + \mu M\Big)| \psi_\text{rot}^M \rangle. \quad &&\text{($L_y$ odd)}
\end{align}\label{eq:alternative-scar-state-energies}%
\end{subequations}
The scar states hence form a tower with equal energy spacing.
For even $L_y$, a degenerate subspace of energy eigenstates resides at each scar energy $E = \mu M$ for $M = -L_x L_y / 2, -L_x L_y / 2 + 2, \ldots, L_x L_y / 2$. 
For the system sizes considered, we numerically find that the subspace at energy $E = \mu M$ consists of energy eigenstates with magnetization $M$.
We further find that the dimension of the subspace at energy $E = \mu M$ is given by $L_x L_y / 2 \choose M / 2 + L_x L_y / 4$ for the system sizes considered.

We also study a related model hosting the same scar states without the degenerate subspaces.
Consider the operator
\begin{align}
	\hat H_\text{nnn} = c'\big(\bm S_{\bm r_1} \cdot \bm S_{\bar{\bm r}_1} - \bm S_{\bm r_2} \cdot \bm S_{\bar{\bm r}_2}\big)
	\label{eq:next-to-nearest-neighbor-term}
\end{align}
where ${\bm r}_1 = (L_x / 2 - 1, L_y / 2 - 1)$, ${\bm r}_2 = (L_x / 2 - 1, L_y / 2)$ are in part $\mathcal A$ and $\bar{\bm r}_1 = (L_x / 2, L_y / 2)$, $\bar{\bm r}_2 = (L_x / 2, L_y / 2 - 1)$ are obtained by rotating ${\bm r}_1$ and ${\bm r}_2$ around $\bm {r}_\text{center}$ by an angle $\pi$.
The operator $\hat H_\text{nnn}$ connects two next-to-nearest neighbor pairs across the boundary of parts $\mathcal A$ and $\mathcal B$.
We study the model 
\begin{equation}
	\hat{H}_\text{rot}' = \hat{H}_\text{rot} + \hat{H}_\text{nnn}
	\label{eq:H_rot'}
\end{equation}
for even $L_y$.
The scar states are eigenstates of $\hat{H}_\text{rot}'$ with energy $\hat{H}_\text{rot}'|\psi_\text{rot}^M\rangle = \mu M|\psi_\text{rot}^M\rangle$.
For generic parameter values, the scar states are the only eigenstates at the energies $E = \mu M$.

\subsection{Localization}
\label{subsec:localization-EPR}
We study the two models from Eqs.~\eqref{eq:parent-Hamiltonian-rotated-EPR-state} and \eqref{eq:H_rot'} at different disorder strengths and investigate to what extent the models localize.
We generally consider parameter values $J = c = c' = \mu = 1$ and the largest magnetization sector $M = 0$ but we expect similar results for other parameter values and symmetry sectors.
For even $L_y$, the Hamiltonian $\hat{H}_\text{rot}$ restricted to the $M = 0$ sector anticommutes with the rotation operator $\hat{\mathcal R}_\pi$.
Consequently, for each eigenstate with energy $E_i$, there exists a different eigenstate with energy $-E_i$ and the spectrum is hence symmetric around $E = 0$.
The spectrum contains a degenerate subspace at energy $E = 0$ which includes a scar state.
Therefore, we generally avoid the center of the spectrum when investigating the localization of generic eigenstates.
Instead, we study eigenstates with energies slightly below the center of the spectrum, i.e., eigenstates with energies close to $E = (6 E_\text{min} + 4 E_\text{max})/10$.

\begin{figure}
\includegraphics[width=\linewidth]{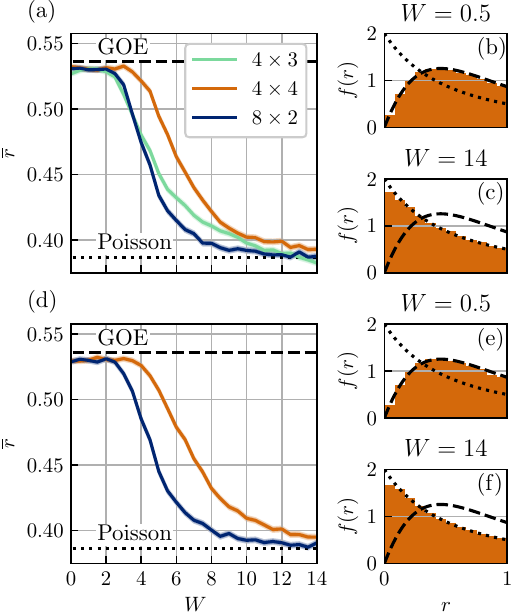}
\caption{
	Level spacing statistics for the Hamiltonians (a)-(c) $\hat{H}_\text{rot}$ and (d)-(f) $\hat{H}_\text{rot}'$.
	We consider parameters $J = c = c' = \mu = 1$ and the symmetry sector $M=0$.
	Panels (a) and (d) display the mean adjacent gap ratio $\overline{r}$ as a function of disorder strength $W$ for different system sizes.
	The adjacent gap ratio is averaged over the $10^2$ energies closest to $(6E_\text{min} + 4E_\text{max}) / 10$ in $10^3$ disorder realizations.
	The shaded areas show two standard deviations on the estimate of the mean.
	The expected value for GOE is shown by the dashed lines and for the Poisson distribution by the dotted lines.
	For both models, the average adjacent gap ratio agrees with GOE at weak disorder indicating that the models are thermal.
	At strong disorder, the average adjacent gap ratio agrees with the Poisson value indicating that both models are localized.
	Panels (b)-(c) and (e)-(f) show the distribution of the adjacent gap ratio for system size $4\times 4$.
	The disorder is weak $W=0.5$ in panels (b) and (e) while it is strong $W = 14$ in panels (c) and (f).
	The distributions $f_\text{GOE}$ and $f_\text{Poisson}$ from Eq.~\eqref{eq:probability-density-functions} are displayed as dashed and dotted curves.
	In both models, the distribution of the adjacent gap ratio agrees with $f_\text{GOE}$ at weak disorder and the distribution agrees with $f_\text{Poisson}$ at strong disorder.
}
\label{fig:adjacent-gap-ratio-alternative}
\end{figure}
We investigate how the models behave with increasing disorder strength.
Figures \ref{fig:adjacent-gap-ratio-alternative}(a) and \ref{fig:adjacent-gap-ratio-alternative}(d) illustrate the mean adjacent gap ratio as a function of disorder strength for respectively $\hat{H}_\text{rot}$ and $\hat{H}_\text{rot}'$. 
For each model, the adjacent gap ratio is averaged over the $10^2$ eigenstates closest in energy to $(6 E_\text{min} + 4 E_\text{max})/10$ in $10^3$ disorder realizations.
At weak disorder, the mean adjacent gap ratio agrees with GOE indicating that the models are thermal.
The mean adjacent gap ratio decreases with increasing disorder strength and coincides with the Poisson value at strong disorder.
This behavior signals that the models localize at strong disorder.
Figures~\ref{fig:adjacent-gap-ratio-alternative}(b)-(c) and \ref{fig:adjacent-gap-ratio-alternative}(e)-(f)  display the distribution of the adjacent gap ratio at weak and strong disorder for $\hat{H}_\text{rot}$ and $\hat{H}_\text{rot}'$.
For both models, the distribution agrees with GOE at weak disorder and it agrees with the Poisson distribution at strong disorder.

\begin{figure*}
	\includegraphics[width=\linewidth]{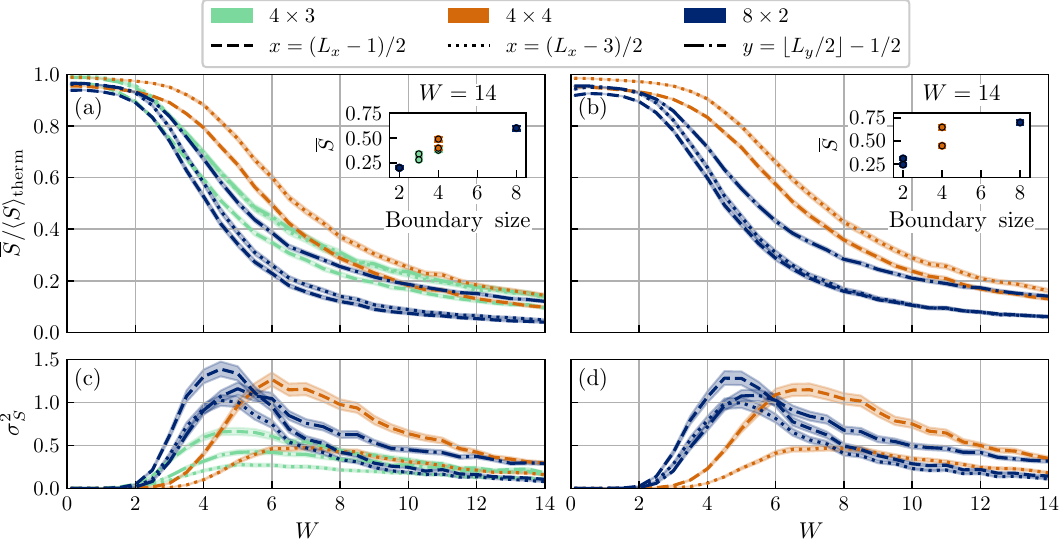}
	\caption{
		(a)-(b) The average entanglement entropy $\overline{S}$ and (c)-(d) the variance of entropy $\sigma_{S}^2$ of the eigenstate with energy closest to $(6E_\text{min} + 4E_\text{max}) / 10$ over $2\times 10^3$ disorder realizations.
		Panels (a) and (c) correspond to the model $\hat{H}_\text{rot}$ while panels (b) and (d) correspond to $\hat{H}_\text{rot}'$.
		The average and variance are shown as a function of disorder strength $W$ for different system sizes and partitions.
		We consider the partitions illustrated in Fig.~\ref{fig:partitions}.
		The system size is indicated by color and the partition by line style.
		The shaded areas display two standard deviations on the estimate of the mean and variance.
		We consider the parameters $J = c = c' = \mu = 1$ and magnetization $M = 0$.
		In both models, the average entropy agrees with the thermal value $\langle S \rangle_\text{therm}$ from Eq.~\eqref{eq:entropy-thermal} at weak disorder and decreases with increasing disorder strength.
		The insets display the average entropy at strong disorder $W = 14$ as a function of boundary size.
		The variance of the entropy displays a peak as the models transition from being thermal to localized.
		}
	\label{fig:entropy-alternative}
\end{figure*}
We further investigate the transition by studying the entanglement entropy as a function of disorder strength as illustrated in Fig.~\ref{fig:entropy-alternative}.
We consider $2\times 10^3$ disorder realizations and compute the entropy of the eigenstate closest in energy to $(6 E_\text{min} + 4 E_\text{max})/10$.
The average and variance of the entanglement entropy are then computed. Figures~\ref{fig:entropy-alternative}(a) and \ref{fig:entropy-alternative}(c) display the results for $\hat{H}_\text{rot}$ while Figs.~\ref{fig:entropy-alternative}(b) and \ref{fig:entropy-alternative}(d) show the results for $\hat{H}_\text{rot}'$.
At weak disorder, the average entanglement entropy agrees with the thermal value $\langle S \rangle_\text{therm}$ from Eq.~\eqref{eq:entropy-thermal} for both models.
As the disorder strength is increased, the average entropy decreases and the variance displays a peak.
At strong disorder, the average entropy is approximately proportional to the boundary size of the partition.
These results support the claim that the two models are thermal at weak disorder and become localized at strong disorder.

\section{Conclusion}\label{sec:conclusion}
We studied a two-dimensional, disordered model hosting a tower of scar states based on the rainbow scar.
At weak disorder, the nonthermal scar states are embedded among thermal eigenstates.
The model transitions from being thermal to localized with increasing disorder strength.
Using a perturbative approach, we demonstrated that the model displays strong localization at large disorder and we confirmed this statement numerically.
Strong localization refers to energy eigenstates having significant support on a small number of basis states.
Furthermore, we provided general guidelines for obtaining strong localization in other scarred models.
We verified that the model transitions from the thermal phase to being localized with increasing disorder strength by studying the level spacing statistics and entanglement entropy.
We showed that the adjacent gap ratio shifts from GOE to the Poisson distribution with increasing disorder strength.
The entanglement entropy displayed volume-law scaling with system size at weak disorder and area-law scaling at strong disorder.
Consequently, the scar states were identified as inverted scars at strong disorder.
We studied the system dynamics from initial states with support in the scar subspace.
The fidelity displayed persistent revivals when the initial state was fully embedded in the scar subspace.
When the initial state had partial support outside the scar subspace, the fidelity displayed revivals with a smaller amplitude.
We demonstrated that the revival amplitude increases with increasing disorder strength and we interpret this result as the localization protecting the scar revivals.
Finally, we constructed two disordered models hosting a tower of scar states with larger entanglement entropy for generic bipartitions.
We demonstrated that these models localize by studying the adjacent gap ratio and the scaling of entanglement entropy as a function of disorder strength.
Hence, the scar states represent a tower of inverted scars in both models.

\begin{acknowledgments}
	This work has been supported by Carlsbergfondet under Grant No.~CF20-0658 and has received funding from the European Research Council (ERC) under the European Union's Horizon 2020 research and innovation program (Grant Agreement No. 101001902).
\end{acknowledgments}

\appendix

\section{Scar states based on the Einstein-Podolsky-Rosen state}
\label{appendix:scar-state-framework}
Scar states and corresponding parent Hamiltonians may be constructed from the Einstein-Podolsky-Rosen (EPR) state.
Let $\mathcal{H}$ be a Hilbert space of the form $\mathcal{H} = \mathcal{H}_{\mathcal{A}} \otimes \mathcal{H}_{\mathcal{B}}$ with $\mathcal{H}_{\mathcal{A}} = \mathcal{H}_{\mathcal{B}}$ and $\text{basis}(\mathcal{H}_{\mathcal{A}})$ a basis for $\mathcal{\mathcal H_{\mathcal A}}$.
The EPR state with respect to $\mathrm{basis}(\mathcal{H}_\mathcal{A})$ is given by
\begin{align}
|\psi_\text{EPR}\rangle = \frac{1}{\sqrt{\mathrm{dim}(\mathcal{H}_{\mathcal{A}})}} \sum_{|\varphi\rangle \in \text{basis}(\mathcal{H}_\mathcal{A})} |\varphi\rangle \otimes |\varphi\rangle.
\end{align}
where $\mathrm{dim}(\mathcal{H}_{\mathcal{A}})$ is the dimension of $\mathcal{H}_{\mathcal{A}}$. 
For any operator $\hat{\mathcal{O}}$ acting within $\mathcal{H}_{\mathcal{A}}$, the EPR state has the special property
\begin{align}
	(\hat{\mathcal{O}} \otimes \hat{\mathds{1}}) |\psi_\text{EPR}\rangle = (\hat{\mathds{1}} \otimes \hat{\mathcal{O}}^T) |\psi_\text{EPR}\rangle
	\label{eq:EPR-property}
\end{align}
where $\hat{\mathcal{O}}^T$ is the transpose of $\hat{\mathcal{O}}$ with respect to $\text{basis}(\mathcal{H}_{\mathcal{A}})$ \cite{Wildeboer2022}.

Based on the EPR state, one may construct other states which inherit a property similar to Eq.~\eqref{eq:EPR-property}.
Let $\hat{U}_{\mathcal{A}}$ and $\hat{U}_{\mathcal{B}}$ be unitary operators acting within respectively $\mathcal{H}_{\mathcal A}$ and $\mathcal{H}_{\mathcal B}$.
Consider the state
\begin{align}
	|\psi_{\hat{U}_{\mathcal{A}}, \hat{U}_{\mathcal{B}}}\rangle = (\hat{U}_{\mathcal A} \otimes \hat{U}_{\mathcal B}) |\psi_\text{EPR}\rangle.
\end{align}
For any operator $\hat{\mathcal{O}}$ acting within $\mathcal{H}_{\mathcal{A}}$, the state fulfills the following property
\begin{subequations}
\begin{align}
	(\hat{\mathcal O} \otimes \hat{\mathds{1}}) |\psi_{\hat{U}_{\mathcal{A}}, \hat{U}_{\mathcal{B}}} \rangle = (\hat{\mathds{1}} \otimes \hat{\mathcal{O}}')|\psi_{\hat{U}_{\mathcal{A}}, \hat{U}_{\mathcal{B}}} \rangle,
\end{align}
with
\begin{align}
	\hat{\mathcal{O}}' = \hat{U}_{\mathcal{B}} \hat{U}_{\mathcal{A}}^T \hat{\mathcal{O}}^T \hat{U}_{\mathcal{A}}^* \hat{U}_{\mathcal{B}}^\dagger
	\label{eq:O-prime}
\end{align}\label{eq:new-scar-property}%
\end{subequations}
where $({\cdots})^*$ is the complex conjugate with respect to $\text{basis}({\mathcal H}_{\mathcal A})$ and $({\cdots})^\dagger$ is the Hermitian conjugate.
This property may be verified by direct calculation.
Following Ref.~\cite{Wildeboer2022}, Eq.~\eqref{eq:new-scar-property} can be utilized to determine parent Hamiltonians for the state $|\psi_{\hat{U}_{\mathcal{A}}, \hat{U}_{\mathcal{B}}}\rangle$.
Consider a general Hamiltonian of the form
\begin{align}
	\hat{H} = \hat{H}_{\mathcal{A}} \otimes \hat{\mathds{1}} + \hat{\mathds{1}} \otimes \hat{H}_{\mathcal{B}} + \sum_i \lambda_i \hat{\mathcal{O}}_{\mathcal{A}}^{(i)} \otimes \hat{\mathcal{O}}_{\mathcal{B}}^{(i)}
\end{align}
where $\hat{H}_\mathcal{A}$, $\hat{\mathcal{O}}_{\mathcal{A}}^{(i)}$ act within $\mathcal{H}_\mathcal{A}$ while $\hat{H}_\mathcal{B}$, $\hat{\mathcal{O}}_{\mathcal{B}}^{(i)}$ act within $\mathcal{H}_\mathcal{B}$ and $\lambda_i \in \mathbb{R}$.
The state $|\psi_{\hat{U}_{\mathcal{A}}, \hat{U}_{\mathcal{B}}}\rangle$ is an eigenstate of $\hat{H}$ with energy $E$ if
\begin{equation}
\begin{split}
	\Big[\hat{\mathds{1}} \otimes \Big(\hat{H}_{\mathcal{A}}' + \hat{H}_{\mathcal{B}} + \sum_i \lambda_i \hat{\mathcal{O}}_{\mathcal{B}}^{(i)}\hat{\mathcal{O}}_{\mathcal{A}}^{(i)}{}'\Big)\Big] |\psi_{\hat{U}_{\mathcal A}, \hat{U}_{\mathcal B}}\rangle \\
	= E|\psi_{\hat{U}_{\mathcal A}, \hat{U}_{\mathcal B}}\rangle
\end{split}
\end{equation}
where $\hat{H}_{\mathcal{A}}'$ and $\hat{\mathcal{O}}_{\mathcal{A}}^{(i)}{}'$ are determined from Eq.~\eqref{eq:O-prime}.
The rainbow scar is described by this framework and corresponds to the choice $\hat{U}_{\mathcal{A}} = \hat{\mathds{1}}$ and $\hat{U}_{\mathcal{B}} = \hat{\mathcal{M}}$ where $\hat{\mathcal M}$ is the mirror operator, i.e., $|\psi_\text{RB}\rangle = |\psi_{\hat{\mathds 1}, \hat{\mathcal M}}\rangle$.
The state from Eq.~\eqref{eq:rotated-scar} corresponds to $\hat{U}_{\mathcal{A}} = \hat{\mathds{1}}$ and $\hat{U}_{\mathcal{B}} = \hat{\mathcal{R}}_{\pi}$ where $\hat{\mathcal{R}}_{\pi}$ is the rotation operator around the center of the lattice by an angle $\pi$, i.e., $|\psi_\text{rot}\rangle = |\psi_{\hat{\mathds 1}, \hat{\mathcal R}_{\pi}}\rangle$.

\section{A perturbative approach to characterizing the localization}\label{sec:appendix}
At strong disorder, the eigenstates of the Hamiltonian in Eq.~\eqref{eq:disorder-highlighted-Hamiltonian} are expected to localize within a vector space $\mathcal V_{\bm D} = \mathrm{span}\{| \bm D, n \rangle | n = 1, 2 \ldots \}$,
\begin{equation}
	| E_{\bm D, m} \rangle \approx \sum_n c_{\bm D; mn} | \bm D, n \rangle.
	\label{eq:eigenstate-expansion-appendix}
\end{equation}
At first glance, when the sum in Eq.~\eqref{eq:eigenstate-expansion-appendix} contains more than one term, the eigenstates seem to have significant support on many basis states.
It turns out, however, that the eigenstates exhibit stronger localization than predicted by Eq.\ \eqref{eq:eigenstate-expansion-appendix}.
The eigenstates generally localize on small subspaces of $\mathcal V_{\bm D}$.
We describe the localization using degenerate Rayleigh-Schrödinger perturbation theory \cite{Sakurai1994}.
For completeness, we briefly review degenerate perturbation theory before applying it to our model.

\subsection{Degenerate perturbation theory}\label{sec:degenerate-perturbation-theory}
We investigate the model at strong disorder.
The Hamiltonian is rewritten to extract the disorder strength
\begin{equation}
	\hat H = \hat H_0 + W \sum_{\bm r \in \mathcal A} h_{\bm r}' \hat D_{\bm r}.
\end{equation}
where $h_{\bm r}'$ are drawn randomly with uniform propability from the interval $[-1, 1]$.
Next, we define $\lambda = 1/W$ and study the related Hamiltonian $\tilde H = \hat H/W$
\begin{equation}
	\tilde H = \lambda \hat H_0 + \sum_{\bm r \in \mathcal A} h_{\bm r}' \hat D_{\bm r}.
\end{equation}
The two Hamiltonians share all eigenstates $|\tilde E_{\bm D, m} \rangle = | E_{\bm D, m} \rangle$ and the energies are related according to $\tilde E_{\bm D, m} = E_{\bm D, m} / W$.
In the following, we focus on a set of eigenstates with identical disorder indices $\{|\tilde E_{\bm D, m} \rangle | m = 1, 2, \ldots \}$ at strong disorder.
In the extreme limit $\lambda = 0$, these eigenstates are exactly degenerate.
At $0 < \lambda$, we expand the energies and eigenstates in powers of $\lambda$
\begin{subequations}
\begin{align}
	\tilde E_{\bm D, m} =& \sum_{\ell = 0}^\infty \lambda^{\ell} \tilde E_{\bm D, m}^{(\ell)} 
	\label{eq:perturbation-expansion-energies}\\
\begin{split}
	|\tilde E_{\bm D, m} \rangle =& \sum_{n} c_{\bm D; mn}^{(0)} | \bm D, n \rangle\\
	&+ \sum_{\ell = 1}^\infty \sum_{\bm D' \neq \bm D} \sum_n \lambda^\ell c_{\bm D'; mn}^{(\ell)} | \bm D', n \rangle
\end{split}\label{eq:perturbation-expansion-eigenstates}
\end{align}\label{eq:perturbation-expansion}%
\end{subequations}
Note that Eq.\ \eqref{eq:perturbation-expansion-eigenstates} is not normalized to simplify notation.

We investigate how the eigenstates localize within the subspaces $\mathcal V_{\bm D}$ by determining the zero-order coefficients $c_{\bm D; mn}^{(0)}$.
For convenience, we collect these coefficients into vectors $\bm c_m^{(0)}$ with $[\bm c_m^{(0)}]_n = c_{\bm D; mn}^{(0)}$.

We substitute Eq.~\eqref{eq:perturbation-expansion} into $\tilde H |\tilde E_{\bm D, m}\rangle = \tilde E_{\bm D, m} | \tilde E_{\bm D, m} \rangle$ and compare similar powers of $\lambda$.
This leads to the familiar result that the first-order correction to the energies $\tilde E_{\bm D, m}^{(1)}$ are the eigenvalues of the matrix
\begin{equation}
	\Big[\hat M^{(1)}\Big]_{nn'} = \langle \bm D, n | \hat H_0 | \bm D, n' \rangle.
	\label{eq:M1}
\end{equation}
The zero-order coefficients $\bm c_{m}^{(0)}$ are the eigenvectors of the same matrix.
If all eigenvalues of $\hat M^{(1)}$ are different, then every coefficient $c_{\bm D; mn}^{(0)}$ is uniquely determined.
In the case where some eigenvalues of $\hat M^{(1)}$ are degenerate, the corresponding zero-order coefficients $\bm c_{m}^{(0)}$ remain undetermined.
Let $\{\bm v_i^{(1)} | i = 1, 2, \ldots\}$ be eigenvectors of $\hat M^{(1)}$ corresponding to the same eigenvalue.
The yet undetermined coefficients are linear combinations of these eigenvectors $\bm c_m^{(0)} \in \mathrm{span}(\{\bm v_i^{(1)} | i = 1, 2, \ldots\})$.
The correct linear combinations are found as the eigenvectors of $\hat M^{(2)}$ with the corresponding eigenvalues being the second-order energy corrections $E_{\bm D, m}^{(2)}$.
This matrix is obtained from the general expression
\begin{equation}
	\hat M^{(\ell)} = \hat P \hat H_0 \hat T^{(\ell - 1)},
	\label{eq:Mell}
\end{equation}
where $\hat P = \sum_n |\bm D, n \rangle \langle \bm D, n|$ is the projection onto $\mathcal V_{\bm D}$.
The matrices $\hat T^{(\ell)}$ are determined from the following recursion
\begin{subequations}
\begin{align}
	\hat T^{(1)} &= \hat {\mathcal E} \hat H_0 \hat P, \\
	\hat T^{(\ell)} &= \hat {\mathcal E} \Big(\hat H_0 \hat T^{(\ell - 1)} - \sum_{\ell' = 1}^{\ell - 1} E_n^{(\ell - \ell')} \hat T^{(\ell')} \Big)
\end{align}
\end{subequations}
with
\begin{equation}
	\hat {\mathcal E} = \sum_{\bm D' \neq \bm D} \sum_{n'} \frac{1}{E_{\bm D, n}^{(0)} - E_{\bm D', n'}^{(0)}} |\bm D', n' \rangle \langle \bm D', n'|.
	\label{eq:E-matrix}
\end{equation}
If the degeneracy persists to second order, we proceed to third order by diagonalizing $\hat M^{(3)}$ in the relevant subspaces and so on.
In this manner, all coefficients $\bm c_m^{(0)}$ can be found by going to high enough order in perturbation theory.

\subsection{Characterizing the localization}
Returning to the model from Eq.~\eqref{eq:disorder-highlighted-Hamiltonian}, we study the energy eigenstates at strong disorder.
We note that $\hat H_0$ contains two-body kinetic terms.
Furthermore, for two basis states belonging to the same magnetization sector and with identical disorder indices $|\bm D, n\rangle$ and $|\bm D, n'\rangle$, the spins are different on more than two sites.
This implies that the offdiagonal elements of $M^{(1)}$ are zero for all $\bm D$.
Hence, for each unique eigenvalue of $M^{(1)}$, the corresponding eigenstate localizes on a single basis state, i.e., $c_{\bm D; mn}^{(0)} = \delta_{mn}$ where $\delta_{mn}$ is the Kronecker delta.
These eigenstates display strong localization. 
Similar arguments are valid for coefficients $c_{\bm D; mn}^{(0)}$ determined in higher order of perturbation theory.
Let $p_{\bm D}$ be the largest integer such that $\langle \bm D, n | \hat H_0^{p_{\bm D}} | \bm D, n' \rangle = 0$ for all $n\neq n'$.
For instance, we have $p_{\bm D} = 4$ in the example from Eq.\ \eqref{eq:introducing-example-action}.
This integer is typically large since $\hat H_0$ only contains local two-body terms.
Inspecting Eqs.~\eqref{eq:Mell}-\eqref{eq:E-matrix}, the offdiagonal elements of $\hat M^{(\ell)}$ vanish for all $\ell \leq p_{\bm D}$.
Hence, if $\bm c_m^{(0)}$ is determined in $\ell$-th order of perturbation theory with $\ell < p_{\bm D}$, then the corresponding energy eigenstate localizes on a single basis state.
From these considerations, we expect most energy eigenstates to localize on smaller subspaces than predicted by Eq.\ \eqref{eq:eigenstate-expansion}.
We remark that the perturbative arguments are not valid for $\mathcal V_{\bm D_0}$ with $[\bm D_0]_{\bm r} = 0$ for all $\bm r \in \mathcal A$.
This subspace is insensitive to disorder, i.e., $\hat D_{\bm r} |\bm D_0, n\rangle = 0$ for all $\bm r \in \mathcal A$.
We expect the eigenstates with significant support in $\mathcal V_{\bm D_0}$ to remain nonlocalized even at strong disorder.

While most energy eigenstates display strong localization, some eigenstates localize in a weaker sense.
These eigenstates have significant support on multiple basis states at strong disorder.
However, the nonlocalized nature of these eigenstates is not visible in the dynamics of observables at small times.
Let $\{|\tilde E_{\bm D, m} \rangle | m = 1, 2, \ldots \}$ be a set of such eigenstates and recall that the degeneracy is lifted in second or higher order of perturbation theory, i.e., $\tilde E_{\bm D, m}^{(1)} = \tilde E_{\bm D, m'}^{(1)}$ for $m \neq m'$.
We return to the energies $E_{\bm D, m}$ and eigenstates $|E_{\bm D, m} \rangle$ of the original Hamiltonian.
Using the relation $E_{\bm D, m} = \lambda \tilde E_{\bm D, m}$, the gap between two such energy eigenstates is given by
\begin{equation}
	E_{\bm D, m} - E_{\bm D, m'} = \lambda (E_{\bm D, m}^{(2)} - E_{\bm D, m'}^{(2)}) + \mathcal O(\lambda^2)
	\label{eq:energy-gap}
\end{equation}
where $\mathcal O(\lambda^2)$ refer to second or higher order terms in $\lambda$.
Equation \eqref{eq:energy-gap} shows that the energy gap vanishes in the limit of very strong disorder.
The fact that these eigenstates have similar energy eliminates all dynamics at small times.
To illustrate this point, consider multiple eigenstates with significant support on the same product states $|E_{\bm D, m} \rangle \approx \sum_{n} c_{\bm D; mn} | \bm D, n \rangle$.
The system is initialized as one of these product states $| \psi(0) \rangle = |\bm D, n \rangle$ and subsequently time evolved $|\psi(t) \rangle = e^{-i \hat H t}|\psi(0)\rangle$.
The time evolved state only differs slightly from the initial state at small times, i.e., $|\langle\psi(0)|\psi(t) \rangle|^2 \approx 1$.

\subsection{Accuracy of the perturbative approach}
\begin{figure}
	\includegraphics[width=\linewidth]{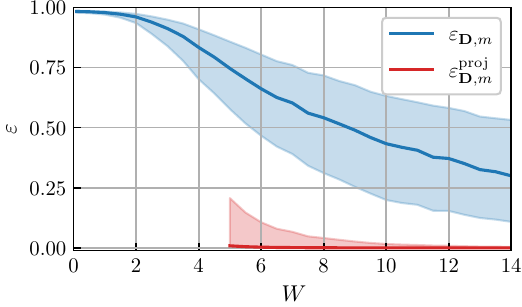}
	\caption{
		The error measures $\varepsilon_{\bm D, m}$ (blue upper curve) and $\varepsilon_{\bm D, m}^\text{proj}$ (red lower curve) as a function of disorder strength $W$ for system size $L_x \times L_y = 4 \times 3$, parameters $J = c = \mu = 1$ and symmetry sector $M = 0$.
		We consider the distribution of the error measures over the full spectrum in $10^3$ disorder realizations.
		The figure displays the median (solid line) and the interquartile range (shaded area) of the distribution of each error measure.
		Note that $\varepsilon_{\bm D, m}^\text{proj}$ can only be determined at strong disorder $5 \leq W$ where the disorder indices of energy eigenstates are well-defined.
		The error $\varepsilon_{\bm D, m}$ decreases with increasing disorder strength but remains finite due to support on other subspaces $\mathcal V_{\bm D'}$ with $\bm D' \neq \bm D$.
		However, the error $\varepsilon_{\bm D, m}$ will vanish at large enough disorder.
		The quantity $\varepsilon_{\bm D, m}^\text{proj}$ is generally close to zero indicating that the zero-order eigenstates correctly describe the spectrum.
	}
	\label{fig:perturbation-plot}
\end{figure}
The analysis above is, strictly speaking, only true in the limit $\lambda \to 0$ or alternatively $W \to \infty$.
However, we expect the results to be a good approximation for large, finite disorder strength.
We verify the formulas presented in Sec.~\ref{sec:degenerate-perturbation-theory} by computing the zero-order coefficients $c_{\bm D;mn}^{(0)}$ using Eqs.\ \eqref{eq:M1}-\eqref{eq:E-matrix}.
The corresponding zero-order eigenstates are denoted by $|E_{\bm D, m}^{(0)} \rangle = \sum_n c_{\bm D;mn}^{(0)} |\bm D, n\rangle$.
Furthermore, we determine the true eigenstates $|E_i\rangle$ at different disorder strengths using exact diagonalization.
For each $|E_{\bm D, m}^{(0)} \rangle$, we determine the most similar exact eigenstate, i.e., $\mathrm{argmax}_{i}\big(|\langle E_{\bm D, m}^{(0)} | E_i \rangle |^2\big)$, and compute the ``error''
\begin{equation}
	\varepsilon_{\bm D, m} = 1 - \max_{i}\big(|\langle E_{\bm D, m}^{(0)} | E_i \rangle |^2\big).
	\label{eq:epsilon}
\end{equation}
The distribution of $\varepsilon_{\bm D, m}$ is determined from $10^3$ disorder realizations.
The error is displayed as a function of disorder strength in Fig.\ \ref{fig:perturbation-plot}.
At weak disorder, the error is near its maximum value $\varepsilon_{\bm D, m} \approx 1$ since all exact eigenstates are delocalized.
The error decreases with increasing disorder strength but remains strictly larger than zero even at very strong disorder.
The error remains finite since all exact eigenstates have a small overlap with other subspaces $\mathcal V_{\bm D'}$ with $\bm D' \neq \bm D$.
We expect the error to vanish for large enough disorder strength.

We further investigate the validity of our approach by considering another error measure.
In the following discussion, we consider sufficiently strong disorder $5 \leq W$ so each exact eigenstate has well-defined disorder indices $\bm D$.
Let $|E_{\bm D, m}^\text{proj} \rangle$ be the normalized projection of $|E_{\bm D, m}\rangle$ onto the subspace $\mathcal V_{\bm D}$
\begin{equation}
	| E_{\bm D, m}^\text{proj} \rangle = \frac{\hat P| E_{\bm D, m} \rangle}{\sqrt{\langle E_{\bm D, m} | \hat P | E_{\bm D, m} \rangle}}
\end{equation}
where $\hat P$ is the projection onto $\mathcal V_{\bm D}$.
We consider the fidelity between $| E_{\bm D, m}^\text{proj} \rangle$ and the matching zero-order eigenstate from perturbation theory $|\langle E_{\bm D, m}^{(0)} | E_{\bm D, m}^\text{proj} \rangle |^2$.
We study the error
\begin{equation}
	\varepsilon_{\bm D, m}^\text{proj} = 1 - |\langle E_{\bm D, m}^{(0)} | E_{\bm D, m}^\text{proj} \rangle |^2.
	\label{eq:epsilon-proj}
\end{equation}
The distribution of $\varepsilon_{\bm D, m}^\text{proj}$ is determined from $10^3$ disorder realizations.
Figure \ref{fig:perturbation-plot} illustrates this error as a function of disorder strength.
The error approaches zero at large disorder indicating that the perturbative approach yields the correct zero-order eigenstates at strong, but finite, disorder strength.

\bibliography{bibliography}

\end{document}